\documentclass[fleqn,usenatbib]{mnras}

\usepackage[T1]{fontenc}
\usepackage{ae,aecompl}

\usepackage{graphicx}	
\usepackage{amsmath}	
\usepackage{amssymb}	

\title[X-ray absorption in the Milky Way using {\tt IONeq}]{Probing the structure of the gas in the Milky Way through X-ray high resolution spectroscopy}

\author[Gatuzz \& Churazov]{
Efra\'in~Gatuzz$^{1}$\thanks{E-mail: egatuzz@mpa-garching.mpg.de}
and Eugene~Churazov$^{1,2}$ 
\\
$^{1}$Max-Planck-Institut f\"ur Astrophysik,
85741 Garching bei M\"unchen, Germany \\
$^{2}$Space Research Institute (IKI), Profsoyuznaya 84/32, Moscow 117997, Russia
}

\date{Accepted XXX. Received YYY; in original form ZZZ}

\pubyear{2016}

\begin{document}
\label{firstpage}
\pagerange{\pageref{firstpage}--\pageref{lastpage}}
\maketitle

\begin{abstract}
We have developed a new X-ray absorption model, called {\tt IONeq}, which computes the optical depth $\tau(E)$ simultaneously for ions of all abundant elements, assuming ionization equilibrium and taking into account turbulent broadening. We use this model to analyze the interstellar medium (ISM) absorption features in the Milky Way for a sample of 18 galactic (LMXBs) and 42 extragalactic sources (mainly Blazars). The absorbing ISM was modeled as a combination of three components/phases - neutral ($T\lesssim1\times10^{4}$ K), warm ($T\sim5\times 10^{4}$ K) and hot ($T\sim2\times10^{6}$ K). We found that the spatial distribution of both, neutral and warm components, are difficult to describe using smooth profiles due to nonuniform distribution of the column densities over the sky. For the hot phase we used a combination of a flattened disk and a halo, finding comparable column densities for both spatial components, in the order of $\sim 6-7\times10^{18}\;{\rm cm^{-2}}$, although this conclusion depends on the adopted parametrization. If the halo component has sub-solar abundance $Z$, then the column density has be scaled up by a factor $\frac{Z_\odot}{Z}$.   The vertically integrated column densities of the disk components suggests the following mass fractions for these three ISM phases in the Galactic disk: neutral $\sim~89\%$, warm $\sim 8\%$ and hot $\sim 3\%$ components, respectively. The constraints on the radial distribution of the halo component of the hot component are weak. 
\end{abstract}




\section{Introduction}\label{sec_intro}
The interstellar medium (ISM) is an important constituent of galaxies, affecting star formation and their overall evolution. This complex environment shows multiple phases characterized by different gas temperatures \citep[e.g.][]{mck77,dra11}. Using the high-resolution X-ray spectroscopy technique the physical conditions that characterize each phase can be analyzed. In order to carry out such an analysis bright X-ray sources with featureless spectra are required, allowing the identification of the absorption features that are imprinted in the spectra by the ISM located on the line of sight between the observer and the source. In this way, X-ray spectra provide information about basic properties of the ISM such as ionization degree, elemental abundances, column densities and temperatures.

A method commonly used to study the absorption lines due to the different ionic species observed in X-ray spectra is through fitting a Gaussian profile to individual lines. In this way the measurement of the equivalent widths (EWs) allows the estimation of the column densities through the curve of growth analysis \citep[e.g.][]{jue04}. This method has been used to study the absorption lines from ionic species such as {\rm O}~{\sc i}, {\rm O}~{\sc ii}, {\rm O}~{\sc iii}, {\rm O}~{\sc vii}, {\rm Ne}~{\sc i}, {\rm Ne}~{\sc ii}, {\rm Ne}~{\sc iii} and {\rm Ne}~{\sc ix} associated to the ISM \citep[e.g.][]{sch02, tak02,jue04,jue06, yao09,lia13,mil13,luo14,mil15,nic16a,nic16c}. An important issue of such analysis is that if the lines are saturated and not resolved there is a degeneracy between the broadening velocity and the column density that can lead to under(over)-estimation of the column densities \citep[e.g.][]{yao09}. If multiple transitions for the same ion are observed, this degeneracy can be broken \citep[e.g.][]{nic16a}.

On the other hand, there have been attempts to model the ISM with physically motivated models, which include ionization equilibrium conditions, to study the physical conditions of the gas directly \citep{pin10,pin13,gat16,nic16a}. As has been shown by \citet{gat16}, the use of physical models helps to prevent misidentification of absorption features compared with traditional phenomenological fitting methods. For instance, using such a physical model  \citet{nic16c}  concluded that the absorption line located at $\sim 22.275$~\AA, which was previously attributed to the warm-hot intergalactic medium (WHIM), is produced by the {\rm O}~{\sc ii} K$\beta$ transition associated with the local ISM. In this sense, a detailed modeling of the neutral, warm and hot absorption features using the most up to date atomic data available is crucial to understand the Milky Way contribution to the atomic and ionic column densities.

In the recent years \citet{gat13a,gat13b,gat14,gat15,gat16} have performed a comprehensive analysis of the galactic ISM using {\it Chandra} and {\it XMM-Newton} X-ray high-resolution spectra, finding an accurate modeling of the neutral, singly and double ionized absorbers. In this paper, we present an extension of such analysis for which we have developed a new X-ray absorption model, called {\tt IONeq}, which includes ionization equilibrium, velocity broadening and predicts the absorption features due to all astrophysically abundant elements. The outline of this paper is as follows. In Section~\ref{sec_ioneq} we describe the {\tt IONeq} X-ray absorption model. In Section~\ref{sec_data} the data selection and the reduction are summarized. Results obtained from the spectral fits are reviewed in Section~\ref{sec_result}, followed by a detailed discussion of the density profiles modeling in Section~\ref{sec_density}. In Section~\ref{sec_previous} we briefly compare our results with previous studies while possible caveats in our analysis are listed in Section~\ref{sec_cave}.  Finally, we summarize the results of our analysis in Section~\ref{sec_con}.

\begin{table*}
\caption{\label{tab1}The sample of Galactic sources.}
\small
\centering
\begin{tabular}{lcccccc}
\hline
Source      & Galactic &Distance& $N({\rm H}) $ - 21~cm & $N({\rm H})$ - Neutral & $N({\rm H})$ - Warm& $N({\rm H})$ - Hot \\
   &coordinates&(kpc)& ($10^{21}$cm$^{-2}$) &  ($10^{21}$cm$^{-2}$) &  ($10^{20}$cm$^{-2}$) & ($10^{19}$cm$^{-2}$) \\
\hline
4U~1254--69	&$(	303.48	;	-6.42	)$&$	13	\pm	3	$ $^{a}$&$	3.46	$&$	3.20	\pm	0.01	$&$	4.96	\pm	0.43	$&$	10.27	\pm	1.50	$\\
4U~1543--62	&$(	321.76	;	-6.34	)$&$	7		$ $^{b}$&$	3.79	$&$	2.37	\pm	0.07	$ &$	3.05	\pm	0.77	$&$	5.43	\pm	2.84	$\\
4U~1636--53	&$(	332.91	;	-4.82	)$&$	6	\pm	0.5	$ $^{c}$&$	4.04	$&$	3.85	\pm	0.10	$&$	4.13	\pm	0.59	$&$	8.34	\pm	3.13	$\\
4U~1735--44	&$(	346.05	;	-6.99	)$&$	9.4	\pm	1.4	$ $^{d}$&$	3.96	$&$	3.65	\pm	0.11	$&$	5.41	\pm	1.33	$&$	9.11	\pm	3.25	$\\
4U~1820--30	&$(	2.79	;	-7.91	)$&$	7.6	\pm	0.4	$ $^{e}$&$	2.33	$&$	1.00	\pm	0.06	$&$	1.14	\pm	0.62	$&$	6.24	\pm	1.26	$\\
4U~1915--05	&$(	31.36	;	-8.46	)$&$	8.8	\pm	1.3	$ $^{d}$&$	3.72	$&$	3.73	\pm	0.21	$&$	2.63	\pm	3.01	$&$	8.03	\pm	6.36	$\\
Aql~X--1 	&$(	35.72	;	-4.14	)$&$	5.2	$ $^{d}$&$	4.30	$&$	3.89	\pm	0.04	$&$	1.36	\pm	0.45	$&$	5.13	\pm	1.60	$\\
Cygnus~X--2	&$(	87.33	;	-11.32	)$&$	13.4	\pm	2	$ $^{d}$&$	3.09	$&$	2.00	\pm	0.02	$&$	1.03	\pm	0.31	$&$	3.09	\pm	1.05	$\\
GRO~J1655--40	&$(	344.98	;	2.46	)$&$	3.2	\pm	0.2	$ $^{d}$&$	7.22	$&$	6.41	\pm	0.05	$&$	6.69	\pm	0.61	$&$	7.35	\pm	1.79	$\\
GS~1826--238	&$(	9.27	;	-6.09	)$&$	6.7		$ $^{c}$&$	3.00	$&$	3.16	\pm	0.04	$&$	1.15	\pm	0.54	$&$	12.35	\pm	4.31	$\\
GX~339--4	&$(	338.94	;	-4.33	)$&$	10	\pm	4.5	$ $^{f}$&$	5.18	$&$	4.71	\pm	0.06	$&$	7.87	\pm	0.64	$&$	11.22	\pm	1.49	$\\
GX~349+2	&$(	349.10	;	2.75	)$&$	9.2		$ $^{g}$&$	6.13	$&$	5.62	\pm	0.16	$&$	7.56	\pm	1.14	$&$	7.52	\pm	1.52	$\\
GX~9+9/4U~1728--16	&$(	8.51	;	9.04	)$&$	4.4	\pm	0	$ $^{g}$&$	3.31	$&$	3.24	\pm	0.06	$&$	2.72	\pm	0.73	$&$	3.42	\pm	1.22	$\\
HETEJ1900.1--2455	&$(	11.30	;	-12.87	)$&$	5		$ $^{f}$&$	1.76	$&$	0.98	\pm	0.02	$&$	-0.03	\pm	0.16^{j}	$&$	1.94	\pm	1.47	$\\
SAX~J1808.4--3658	&$(	355.39	;	-8.15	)$&$	2.8		$ $^{c}$&$	1.76	$&$	0.99	\pm	0.08	$&$	1.49	\pm	0.11	$&$	4.36	\pm	0.87	$\\
Ser~X--1	&$(	36.12	;	4.84	)$&$	11.1	\pm	1.6	$ $^{d}$&$	5.42	$&$	4.39	\pm	0.07	$&$	5.33	\pm	0.87	$&$	5.39	\pm	1.58	$\\
Swift~J1753.5--0127	&$(	24.90	;	12.19	)$&$	5.4		$ $^{h}$&$	2.98	$&$	1.06	\pm	0.01	$&$	-0.64	\pm	0.03^{j}	$&$	3.23	\pm	0.56	$\\
XTE~J1817--330	&$(	359.82	;	-8.00	)$&$	2.5	\pm	1.5	$ $^{i}$&$	2.29	$&$	2.12	\pm	0.03	$&$	3.37	\pm	0.35	$&$	5.58	\pm	0.90	$\\
\hline
\multicolumn{7}{p{15cm}}{21~cm measurements are taken from \citet{wil13}. $N({\rm H})$-Neutral, $N({\rm H})$-Warm and $N({\rm H})$-Hot values are obtained from the best-fit described in Section~\ref{sec_data}. Distances are taken from: $^a$\citet{int03}; $^{b}$\citet{wan04}; $^c$\citet{gall08}; $^d$\citet{jon04}; $^e$\citet{kul03}; $^f$\citet{hyn04}; $^g$\citet{gri02}; $^h$\citet{kaj16}; and $^i$\citet{sal06}. $^j$ - in these sources the {\rm O}~{\sc ii} $K_\alpha$ line is strongly affected by RGS instrumental feature. }
\end{tabular}
\end{table*}


\begin{table*}
\caption{\label{tab2}The sample of extragalactic sources.}
\small
\centering
\begin{tabular}{lccccccc}
\hline
Source      & Galactic & $N({\rm H}) $ - 21~cm   & $N({\rm H})$ - Neutral & $N({\rm H})$ - Warm & $N({\rm H})$ - Hot \\
   &coordinates &($10^{20}$cm$^{-2}$) & ($10^{20}$cm$^{-2}$) &  ($10^{19}$cm$^{-2}$) & ($10^{19}$cm$^{-2}$) \\
\hline
1ES~1028+511	&$(	161.44	;	54.44	)$&$	1.26	$&$	2.30	\pm	0.78	$&$	-1.96	\pm	1.35	$&$	1.42	\pm	1.77	$\\
1ES~1553+113	&$(	21.91	;	43.96	)$&$	4.35	$&$	3.92	\pm	0.40	$&$	4.18	\pm	2.26	$&$	0.90	\pm	0.77	$\\
1ES~1927+654	&$(	96.98	;	20.96	)$&$	9.20	$&$	9.19	\pm	1.57	$&$	-6.31	\pm	2.22	$&$	3.27	\pm	4.10	$\\
1ES~0120+340	&$(	130.34	;	-28.06	)$&$	5.23	$&$	5.56	\pm	1.42	$&$	10.41	\pm	9.82	$&$	-0.82	\pm	0.77	$\\
1H~0414+009	&$(	191.81	;	-33.16	)$&$	13.70	$&$	9.49	\pm	1.19	$&$	5.15	\pm	7.46	$&$	1.54	\pm	2.53	$\\
1H~0707--495	&$(	260.17	;	-17.67	)$&$	6.55	$&$	5.77	\pm	0.38	$&$	4.15	\pm	2.51	$&$	1.62	\pm	0.88	$\\
1H~1426+428	&$(	77.49	;	64.90	)$&$	1.14	$&$	1.76	\pm	1.04	$&$	0.20	\pm	0.81	$&$	1.74	\pm	1.54	$\\
3C~120	&$(	190.37	;	-27.40	)$&$	19.30	$&$	12.90	\pm	0.47	$&$	-1.35	\pm	2.09	$&$	2.03	\pm	1.50	$\\
3C~273	&$(	289.95	;	64.36	)$&$	1.78	$&$	2.55	\pm	0.21	$&$	1.83	\pm	1.31	$&$	4.30	\pm	1.01	$\\
3C~279	&$(	305.10	;	57.06	)$&$	2.22	$&$	4.29	\pm	2.60	$&$	0.00	\pm	6.40	$&$	4.19	\pm	4.26	$\\
3C~382	&$(	61.31	;	17.45	)$&$	9.27	$&$	8.53	\pm	1.51	$&$	5.91	\pm	7.39	$&$	6.30	\pm	4.51	$\\
3C~390.3	&$(	111.44	;	27.07	)$&$	4.51	$&$	4.01	\pm	0.58	$&$	-3.10	\pm	1.98	$&$	3.96	\pm	2.86	$\\
3C~59	&$(	142.04	;	-30.54	)$&$	6.89	$&$	3.57	\pm	4.10	$&$	2.08	\pm	2.84	$&$	2.98	\pm	2.08	$\\
Ark~564	&$(	92.14	;	-25.34	)$&$	6.74	$&$	5.18	\pm	0.32	$&$	9.20	\pm	2.65	$&$	1.08	\pm	0.54	$\\
B0502+675	&$(	143.79	;	15.89	)$&$	14.50	$&$	9.78	\pm	0.99	$&$	12.67	\pm	7.38	$&$	4.10	\pm	2.90	$\\
ESO~141--G055	&$(	338.18	;	-26.71	)$&$	6.41	$&$	6.29	\pm	0.41	$&$	0.69	\pm	2.80	$&$	4.39	\pm	1.91	$\\
ESO~198--G24	&$(	271.64	;	-57.95	)$&$	3.27	$&$	4.29	\pm	0.98	$&$	2.76	\pm	5.56	$&$	6.04	\pm	4.33	$\\
Fairall~9	&$(	295.07	;	-57.83	)$&$	3.43	$&$	5.21	\pm	0.42	$&$	-2.76	\pm	1.41	$&$	1.86	\pm	1.49	$\\
H2356--309	&$(	12.84	;	-78.04	)$&$	1.48	$&$	2.49	\pm	0.42	$&$	2.44	\pm	3.08	$&$	1.34	\pm	1.12	$\\
IRAS13349+2438	&$(	20.60	;	79.32	)$&$	1.07	$&$	-1.18	\pm	0.51	$&$	2.21	\pm	5.75	$&$	-0.12	\pm	1.98	$\\
IZw1	&$(	123.75	;	-50.17	)$&$	6.01	$&$	6.38	\pm	2.08	$&$	0.00	\pm	4.13	$&$	0.86	\pm	2.49	$\\
MR2251--178	&$(	46.20	;	-61.33	)$&$	2.67	$&$	1.29	\pm	0.34	$&$	-0.54	\pm	2.41	$&$	1.75	\pm	1.94	$\\
Mrk~1044	&$(	179.69	;	-60.48	)$&$	3.88	$&$	1.40	\pm	1.08	$&$	-2.18	\pm	4.32	$&$	5.37	\pm	4.60	$\\
Mrk~110	&$(	165.01	;	44.36	)$&$	1.39	$&$	3.37	\pm	0.92	$&$	-0.74	\pm	2.83	$&$	1.44	\pm	2.03	$\\
Mrk~279	&$(	115.04	;	46.86	)$&$	1.72	$&$	2.72	\pm	0.52	$&$	6.30	\pm	4.17	$&$	1.58	\pm	1.14	$\\
Mrk~290	&$(	91.49	;	47.95	)$&$	1.75	$&$	2.63	\pm	1.31	$&$	9.07	\pm	9.29	$&$	4.71	\pm	2.84	$\\
Mrk~421	&$(	179.83	;	65.03	)$&$	2.01	$&$	2.53	\pm	0.09	$&$	3.65	\pm	0.82	$&$	1.78	\pm	0.24	$\\
Mrk~501	&$(	63.60	;	38.86	)$&$	1.66	$&$	2.86	\pm	0.37	$&$	2.46	\pm	2.35	$&$	2.86	\pm	1.19	$\\
Mrk~509	&$(	35.97	;	-29.86	)$&$	5.04	$&$	3.59	\pm	0.38	$&$	5.90	\pm	2.93	$&$	3.71	\pm	1.11	$\\
Mrk~841	&$(	11.21	;	54.63	)$&$	2.43	$&$	1.19	\pm	0.66	$&$	5.67	\pm	6.06	$&$	3.10	\pm	2.56	$\\
NGC~3783	&$(	287.46	;	22.95	)$&$	13.80	$&$	10.34	\pm	0.82	$&$	2.40	\pm	3.26	$&$	2.56	\pm	1.73	$\\
NGC~4593	&$(	297.48	;	57.40	)$&$	2.04	$&$	2.95	\pm	0.74	$&$	1.83	\pm	3.28	$&$	3.21	\pm	2.11	$\\
NGC~5548	&$(	31.96	;	70.50	)$&$	1.69	$&$	2.30	\pm	0.65	$&$	9.04	\pm	5.88	$&$	1.02	\pm	1.31	$\\
NGC~7213	&$(	349.59	;	-52.58	)$&$	1.12	$&$	2.04	\pm	0.78	$&$	-1.06	\pm	3.26	$&$	-1.35	\pm	1.13	$\\
NGC~7469	&$(	83.10	;	-45.47	)$&$	5.24	$&$	4.62	\pm	0.56	$&$	4.34	\pm	3.62	$&$	1.73	\pm	1.15	$\\
PG1116+215/Ton~1388	&$(	223.36	;	68.21	)$&$	1.43	$&$	2.46	\pm	1.16	$&$	7.22	\pm	7.43	$&$	0.88	\pm	1.93	$\\
PG1211+143	&$(	267.55	;	74.32	)$&$	3.02	$&$	2.42	\pm	0.85	$&$	-0.65	\pm	4.10	$&$	3.49	\pm	2.65	$\\
PKS~0548--32	&$(	237.57	;	-26.14	)$&$	2.87	$&$	3.74	\pm	0.89	$&$	7.85	\pm	6.34	$&$	-0.20	\pm	1.10	$\\
PKS~0558--504	&$(	257.96	;	-28.57	)$&$	4.18	$&$	2.24	\pm	0.23	$&$	0.68	\pm	1.47	$&$	1.43	\pm	0.66	$\\
PKS2~005--489	&$(	350.37	;	-32.60	)$&$	4.66	$&$	3.59	\pm	1.07	$&$	10.65	\pm	9.18	$&$	3.02	\pm	2.50	$\\
PKS~2155--304	&$(	17.73	;	-52.25	)$&$	1.63	$&$	2.53	\pm	0.10	$&$	1.26	\pm	0.58	$&$	1.62	\pm	0.26	$\\
Tons~180	&$(	138.99	;	-85.07	)$&$	1.54	$&$	1.71	\pm	0.54	$&$	-2.40	\pm	2.08	$&$	1.70	\pm	1.82	$\\
\hline
\multicolumn{7}{p{14.2cm}}{21~cm measurements are taken from \citet{wil13}. $N({\rm H})$-Neutral, $N({\rm H})$-Warm and $N({\rm H})$-Hot values are obtained from the best-fit described in Section~\ref{sec_data}.}
\end{tabular}
\end{table*} 

   \begin{figure}
\includegraphics[width=8.7cm, height=5.0cm]{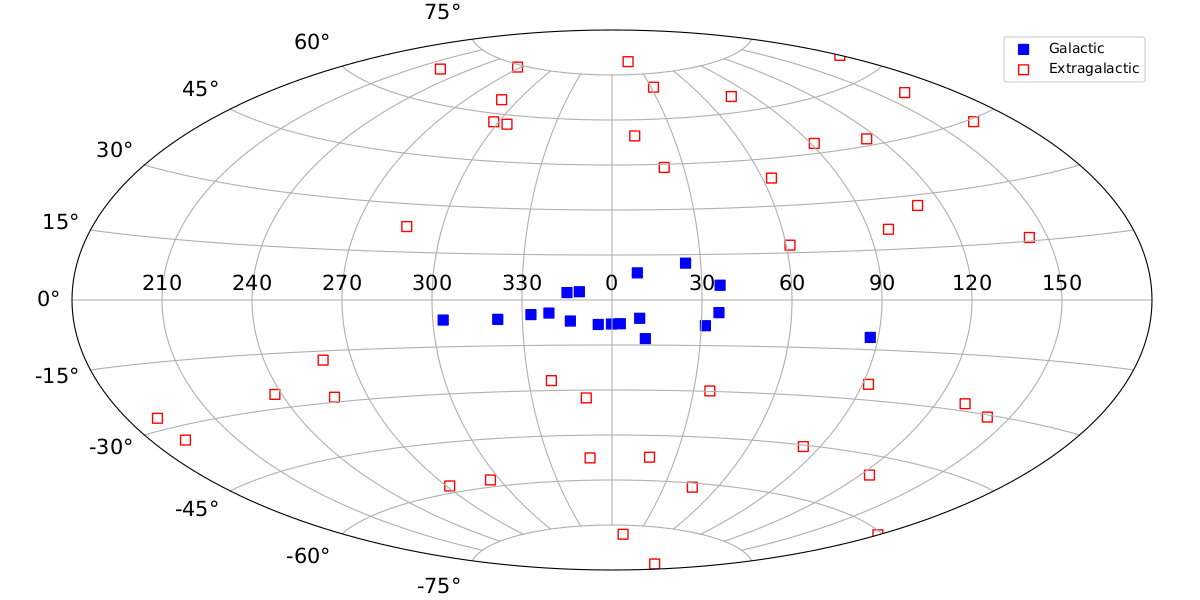}
      \caption{Location map of the X-ray sources in the sample.  }\label{fig_aitoff}
   \end{figure}

\section{Ionization Equilibrium Model: {\tt IONeq}}\label{sec_ioneq}
We have developed a new X-ray absorption model called {\tt IONeq}, which combines the information on the absorption features for astrophysically abundant elements and calculates the ionization balance for them. The latter part is based on the model developed in \citet{chu01}. 
The model assumes ionization equilibrium:
\begin{align}
 &&0=\sum_{j'\neq j}n_{i,j'}R_{i,j'\rightarrow j} - n_{i,j}\sum_{j'\neq j} R_{i,j\rightarrow j'},
\end{align}
where $n_{i,j}$ is the number density of ions $j$ of an element $i$ and the two terms on the r.h.s. represent creation and destruction rates of this ion. $R_{i,j'\rightarrow j}$ is the ionization/recombination rate (per ion $j$), leading to transition from ion $j$ to $j'$. The physical processes included in the creation and destruction rates are photoionization, Auger ionization, direct collisional ionization, radiative recombination and dielectronic recombination.  

We include the photoionization cross sections from \citet{ver96b}, which are convolved with the radiation field to obtain the photoionization rates $R_{i,j}^{ph}$  according to the equation
\begin{align}
&&R_{i,j}^{ph}=\sum_{s} 4\pi\int_{E_{0}}^{\infty}J(E)\sigma_{i,j,s}(E)\frac{dE}{E}
\end{align}
where $E_{0}$ is the ionization potential, $J(E)$ is the ionizing spectrum (in units of erg s$^{-1}$ cm$^{-2}$ sr$^{-1}$ keV$^{-1}$) and $\sigma_{i,j,s}(E)$ is the photoionization cross-section for the shell  $s$ . Ionization of electrons from inner shells can lead to ejection of one or more electrons from higher shells, which are less bounded. This process, known as Auger effect, was included by using Auger probabilities of the various possible final ionization states from  \citet{kas93}. 

In the current {\tt IONeq} version the radiation field $J(E)$ is assumed to have a  power law with the photon-index $\Gamma = 2$, although future versions of the model will allow the use of different ionizing radiation fields. The model includes the ionization parameter  $\xi$ that regulates the photoionization rates and is defined as
\begin{align}\label{eq1}
&&\xi =(4\pi)^{2}\int  \frac{J(E)}{n}dE,
\end{align}
 $n$ is the hydrogen density \citep[see][for definitions of the ionization parameter]{tar69}. The range of energies used in Equation~\ref{eq1} is from 1 Rydberg to 1000 Rydberg, as defined in {\sc xstar} \citep{kal01}.

 For the collisional ionization processes, we include the ionization rates from \citet{vor97}. Radiative recombination rates were taken from \citet{ver96a}. In the case of dielectronic recombination we use the atomic data from \citet{arn85}. Elements up to zinc are incorporated, with their respective ions. It is important to emphasize that {\tt IONeq} does not solve the thermal equilibrium equations, that is the balance
 heating rate$=$cooling rate, from which one obtain the gas temperature. Instead, we treat both, the gas temperature $T$ and the ionization parameter $\xi$, as independent (free) parameters of the model. 

Once the ion fractions are known they can be used to compute the total optical depth $\tau$ defined as
\begin{align}
&&\tau(E) = N({\rm H})\sum_{i,j} A_{i}\xi_{i,j}\sigma_{i,j}(E)\ ,
\label{eq:tau}
\end{align}
where $A_{i}$ is the abundance of the $i$-element with respect to hydrogen, $\xi_{i,j}=n_{i,j}/\sum_{j}n_{i,j}$ is the ion fraction, $\sigma_{i,j}(E)$ is the photoabsorption cross-section of the ion (including photoionization) and $N({\rm H})$ is the hydrogen column density. We use the standard solar abundances from \citet{gre98} and the photoabsorption cross-section used by {\sc xstar}.

We included broadening effects in the expression for $\tau(E)$, assuming that the broadening consists of two components: $v_{th}$ and $v_{turb}$. The former is due to the thermal velocities of ions
\begin{align}
&&v_{th}=\sqrt{\frac{kT}{A_{i}m_{p}}} \ ,
\label{eq:tmax}
\end{align}
where $k$ is the Boltzmann constant, $A_{i}$ is the atomic mass number and $m_{p}$ is the proton mass, while the latter component $v_{turb}$, accounts for non-thermal effects such as turbulent motions in the gas. 

The broadening associated with $v_{turb}$ does not depend on the mass of the ion and the convolution of the optical depth $\tau(E)$ with the corresponding Gaussian is done on-fly when fitting the data. $v_{turb}$ is a free parameter of the model.

The value of $v_{th}$, on the other hand, depends on the gas temperature and the ion mass. We provide two versions of the  {\tt IONeq} model. For one version the value of $v_{th}$ was pre-computed for each ion by setting $T$ in Equation~(\ref{eq:tmax}) to the temperature at which the ion fraction $\xi_{i,j}$ is maximal in CIE.  The  cross sections in Equation~(\ref{eq:tau}) for individual elements/ions were convolved with the corresponding Gaussians and stored in the model. This was done in order to avoid the need to convolve the optical depths for individual ions when fitting the data with the {\tt IONeq} model. For the collisional ionization equilibrium this approximation is reasonable for most of  ions, although for He-like and Ne-like ions, which are present over very broad range of temperatures, a more accurate approach would be to convolve the optical depth for each ion using the value of $T$ provided as a parameter of the model. However, for the purpose of this study the main goal is not to correctly split the line broadening into thermal and turbulent components, but rather have reasonable overall broadening $v^2=v^2_{th}+v^2_{turb}$. As we discuss in the next section we adjust $v_{turb}$ to get the right line ratios for the brightest objects in our sample and then use these values for fainter objects in our sample, assuming that the line broadening is the same for all objects. The second version of the {\tt IONeq} model performs the proper optical depth convolution for each ion using the $T$ parameter.

Finally, the total photoabsorption model takes the form
\begin{align}
&&I_{\rm obs}(E)=I_{\rm source}(E)\exp(-\tau(E))
\end{align}
where $I_{\rm obs}(E)$ is the observed spectrum, $I_{\rm source}(E)$ the spectrum emitted by the source, and $\exp(-\tau(E))$ is the absorption coefficient, which is calculated by the {\tt IONeq} model.    

In summary, the parameters of the model include the temperature T$_{e}$, ionization parameter $\log(\xi)$, hydrogen column density $N({\rm H})$, turbulence velocity $v_{turb}$ and redshift $z$. {\tt IONeq} can be downloaded\footnote{\url{https://heasarc.gsfc.nasa.gov/docs/xanadu/xspec/models/ioneq.html}} for use in the X-ray data analysis package {\sc xspec} \citep{arn96}.

\section{Data sample and fitting procedure}\label{sec_data}

We have compiled a sample of galactic (LMXBs) and extragalactic X-ray sources from both the {\it XMM-Newton} Science Archive (XSA\footnote{\url{http://xmm.esac.esa.int/xsa/}}) and the {\it Chandra} Source Catalog (CSC\footnote{\url{http://cxc.harvard.edu/csc/}}) in order to study the X-ray absorption contribution from various components of the galactic ISM. In order to build our sample we have selected those sources for which we have at least 1000 counts in the oxygen edge absorption region (21--24 \AA). In order to get an unbiased sample we did not impose additional constraints such as, for example, significant detection of a particular line (e.g. {\rm O}~{\sc vii} K$\alpha$).

A total of 18 galactic and 42 extragalactic sources were selected. Tables~\ref{tab1} and~\ref{tab2} show the specifications of the sources, including the galactic coordinates, distances (in the case of LMXBs) and the 21~cm hydrogen column density measurements obtained by \citet{wil13}. Figure~\ref{fig_aitoff} shows an Aitoff projection of the location of the sources in galactic coordinates. It is clear that our final sample allows the analysis of X-ray absorption along multiple lines of sight, including regions in the galactic disk and also in the galactic halo. 

 A total of 257 {\it XMM-Newton} observations were reduced following the standard Scientific Analysis System (SAS, version 16.0.0) threads\footnote{\url{http://xmm.esac.esa.int/sas/current/documentation/threads/}}, while 165 {\it Chandra} observations were reduced following the standard Chandra Interactive Analysis of Observations (CIAO, version 4.9) threads\footnote{\url{http://cxc.harvard.edu/ciao/threads/gspec.html}}. In the case of {\it Chandra} observations we used the {\tt findzo} algorithm\footnote{\url{http://space.mit.edu/cxc/analysis/findzo/}} to estimate the zero order position for the spectra. The spectral fitting was carried out with the {\sc xspec} software (version 12.9.0s\footnote{\url{https://heasarc.gsfc.nasa.gov/xanadu/xspec/}}). Finally, we use the $\chi^{2}$ statistics in combination with the \citet{chu96} weighting method.

           \begin{figure*}
\includegraphics[scale=0.38]{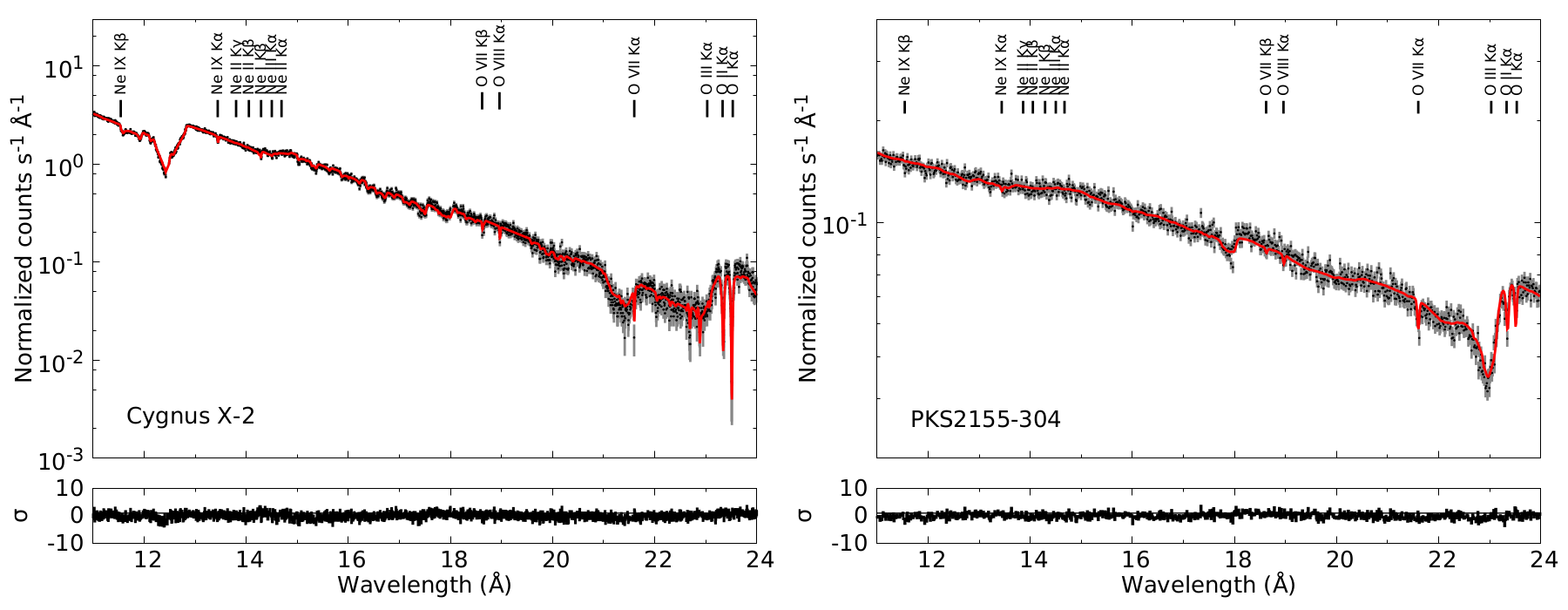}
      \caption{Cygnus~X--2 and PKS2155-304 high-resolution {\it Chandra} spectra. For each source all observations has been combined for illustrative purposes. Solid red line corresponds to the best-fit obtained with the {\tt IONeq(neutral)*IONeq(warm)*IONeq(hot)*Bknpower} model. Absorption lines identified in the Ne and O absorption edge regions are indicated.   }\label{fig_data}
   \end{figure*} 
   
For the purpose of this study we choose to model the absorbing ISM as a combination of three components/phases - neutral (T$_{e}(neutral)\lesssim1\times10^{4}$); this component includes all neutral and molecular gas, warm (T$_{e}(warm)\sim {\rm few} \times 10^{4}$ K) and hot T$_{e}(hot)\sim10^{6}$ K. We have carried out a fit in the 11--24~\AA\ wavelength region for each source listed in Tables~\ref{tab1} and~\ref{tab2} using a simple {\tt IONeq(neutral)*IONeq(warm)*IONeq(hot)*Bknpower} model. The broken power-law component provides a better fit, in general, than a simple powerlaw. For every source all individual observations were fitted simultaneously and the normalization was allowed to vary between them to account for possible variations of the source flux. For each {\tt IONeq} component we assume collisional ionization equilibrium (i.e. $\xi$ $=0$). The value of T$_{e}$ is the parameter that determines the ionic fractions, while $N({\rm H})$ and $v_{turb}$ affects the line equivalent widths. 

Many objects in our sample are too faint to allow accurate determination of both T$_{e}$ and $v_{turb}$. To mitigate this problem we decided to fix these two parameters at the best-fitting values for the brightest sources in the sample. To this end, we use Cygnus~X--2 for the galactic sample and PKS~2155--304 for the extragalactic sample. Figure~\ref{fig_data} shows the best-fitting {\tt IONeq} model. For each source all observations were combined for illustrative purposes. In this plot, black data points correspond to the observations while the
solid red lines represent the best-fit model for each case. The bottom panels show the fit residuals in units of $\sigma$. For the neutral component we set $\log($T$_{e})=4.0$. Our best fit parameters for other two components are $\log($T$_{e})=4.7$ (for the warm component) and $\log($T$_{e})=6.3$ (for the hot component). Table~\ref{tab_ionic} lists the ionic fractions for O and Ne corresponding to these components. In the case of unresolved and saturated lines, there is a degeneracy between the velocity broadening (i.e. Doppler parameter $b=\sqrt{2}v_{turb}$) and the column density, which can leads to under(over)estimation of the column densities \citep{dra11,nic16a}. However, if the spectrum shows absorption lines of the same ion, this degeneracy can be broken. The same approach can be extended to lines of different ions or even different elements, although in this case the results become sensitive to the assumed ionization state of the gas and the abundance ratios. This approach has been used in {\tt IONeq} which fits simultaneously all lines. Later in this section we compare the broadening predicted by  {\tt IONeq} for the hot component with the curve of growth analysis.

Figure~\ref{fig_data} shows the spectra of the brightest sources in our sample. They contain several lines characteristic for each ISM component, allowing one to estimate the broadening. For example, for the hot component the absorption lines from  both {\rm O}~{\sc vii} K$\alpha$ and {\rm O}~{\sc vii} K$\beta$ are dominating, providing almost model independent constraints on the broadening. The {\tt IONeq} best fit parameters for the turbulence broadening are $v_{turb}=75^{+15}_{-55}$ km/s (for neutral and warm components) and $v_{turb}=60^{+28}_{-9}$ km/s (for the hot component) for Cygnus~X--2. In the case of PKS~2155--304 we found $v_{turb}=50^{+12}_{-37}$ km/s (for neutral and warm components) and $v_{turb}=110^{+33}_{-15}$ km/s (for the hot component). Note that the errors quoted above reflect only the photon counting noise and do not include possible systematic errors. 

        \begin{figure}
          \begin{center}
\includegraphics[width=8.5cm,height=8cm]{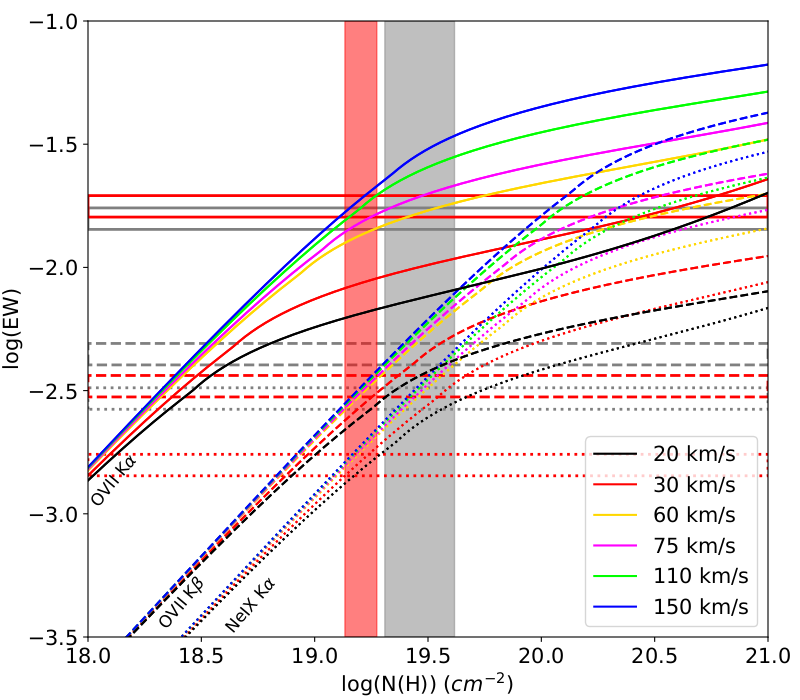}
      \caption{Curve of growths for {\rm O}~{\sc vii} K$\alpha$ (solid lines), {\rm O}~{\sc vii} K$\beta$ (dashed lines) and {\rm Ne}~{\sc ix} K$\alpha$ (doted lines). The color of the lines indicate the broadening velocity.  Vertical shaded regions indicate the hot column densities obtained from the {\tt IONeq} best fit for Cygnus~X-2 (gray) and PKS~2155--304 (red). Pairs of horizontal lines show the uncertainty of measured EWs for these three lines (see text for details). }\label{fig_curve_growth}
      \end{center}
   \end{figure} 
   
The degeneracy between $v_{turb}$ and $N({\rm H})$ for saturated lines leads to large uncertainties in the $v_{turb}$ values. Figure~\ref{fig_curve_growth} shows curves of growth for the hot component absorption lines (solid lines for {\rm O}~{\sc vii} K$\alpha$, dashed lines for {\rm O}~{\sc vii} K$\beta$ and doted lines for {\rm Ne}~{\sc ix} K$\alpha$). The ionic column densities have been rescaled to equivalent hydrogen column densities assuming solar abundances and CIE for T$_{e}=2\times 10^{6}$ K. The vertical shaded regions indicate the $N({\rm H})$ values obtained from the {\tt IONeq} best fit for Cygnus~X-2 (gray) and PKS~2155--304 (red). As the curve of growths show, if only one line is analyzed (e.g. {\rm O}~{\sc vii} K$\alpha$) the $v_{turb}$ and column densities will be essentially unconstrained, however, a multiple lines analysis leads to a better determination of both parameters. We also have measured EWs of individual lines in the spectra of both sources by fitting them with Gaussians and the obtained values, including their uncertainties, are indicated as horizontal regions (solid lines for {\rm O}~{\sc vii} K$\alpha$, dashed lines for {\rm O}~{\sc vii} K$\beta$ and doted lines for {\rm Ne}~{\sc ix} K$\alpha$). We found that the column densities derived from the curve of growth method are in good agreement with the values obtained from the {\tt IONeq} fit, which uses $v_{turb}$ and $N({\rm H})$ as model parameters. 

Table~\ref{tab_vturb} shows a comparison between different $v_{turb}$ values assumed in previous works for the neutral, warm and hot components. Due to the degeneracy between $v_{turb}$ and column densities in the saturated line regime it is difficult to estimate uncertainties for the $v_{turb}$, specially for sources for which multiple transitions for the same ion can not be identified. In this sense, the results listed in Table~\ref{tab_vturb} agree within a factor of $\sim2$, which  might lead to substantial changes in the derived column densities. We decide to use our best-fit parameters for Cygnus~X--2 as reference for the LMXBs and PKS~2155--304 for the extragalactic sources by fixing T$_{e}$ and $v_{turb}$ to the values obtained from the {\tt IONeq} modeling. This essentially means that only the column densities of each ISM component and the broken power law continuum are the remaining free parameters  of the model.  It is important to note that this is the first analysis when all three ISM components are being fitted simultaneously using both {\it Chandra} and {\it XMM-Newton} high-resolution X-ray spectra.

\begin{table}
\caption{\label{tab_vturb}Comparison of broadening velocities ($v_{turb}=b/\sqrt{2}$) adopted in different studies, where $b$ is the Doppler broadening parameter.}
\scriptsize
\centering
\begin{tabular}{lcccc}
\hline
Reference&\multicolumn{2}{c}{Neutral-Warm}&\multicolumn{2}{c}{Hot}\\
&Gal.&Ex.&Gal.&Ex.\\
\hline
This Work&$ 75^{+15}_{-55}$ &$ 50^{+12}_{-37}$  & $ 60^{+28}_{-9}$&$ 110^{+33}_{-15}$\\
\citet{jue04,jue06}&$\leq$ 200&  & & \\
\citet{gup12}&&  & &$\leq$ 141 \\
\citet{mil13}&  & & &106 \\
\citet{pin13}&10 &  &100 & \\
\citet{mil15}& & &  & 106\\
\citet{fan15}& & &  &70\\
\citet{hod16}& & &  &$\leq$ 200\\
\citet{nic16a}&120$\pm$40 &35$\pm$10 & \\
\citet{nic16c}&  &&88 &67 \\
\citet{fae17}& &  & &70\\
\hline
\multicolumn{5}{l}{Velocity in units of km/s. Gal=Galactic sources, Ex=Extragalactic sources}\\
\end{tabular}
\end{table}

\begin{table}
\caption{\label{tab_ionic}Ionic fractions for the neutral,warm and hot components.}
\small
\centering
\begin{tabular}{lccc}
\hline
Ion &\multicolumn{3}{c}{Ion fraction }\\
&Neutral&Warm&Hot\\
&$\log($T$_{e})\lesssim 4.00$&$\log($T$_{e})=4.70$&$\log($T$_{e})=6.30$\\
\hline
{\rm O}~{\sc i}&1.00&0.00&0.00\\
{\rm O}~{\sc ii}&0.00&0.75&0.00\\
{\rm O}~{\sc iii}&0.00&0.25&0.00\\
{\rm O}~{\sc vii}&0.00&0.00&0.71\\
{\rm O}~{\sc viii}&0.00&0.00&0.29\\
{\rm Ne}~{\sc i}&1.00&0.00&0.00\\
{\rm Ne}~{\sc ii}&0.00&0.68&0.00\\
{\rm Ne}~{\sc iii}&0.00&0.32&0.00\\
{\rm Ne}~{\sc ix}&0.00&0.00&0.98\\
{\rm Ne}~{\sc x}&0.00&0.00&0.02\\
\hline
\end{tabular}
\end{table} 
\section{Neutral, warm and hot components in the local ISM}\label{sec_result}
    
      \begin{figure*}
\includegraphics[width=18cm,height=18cm]{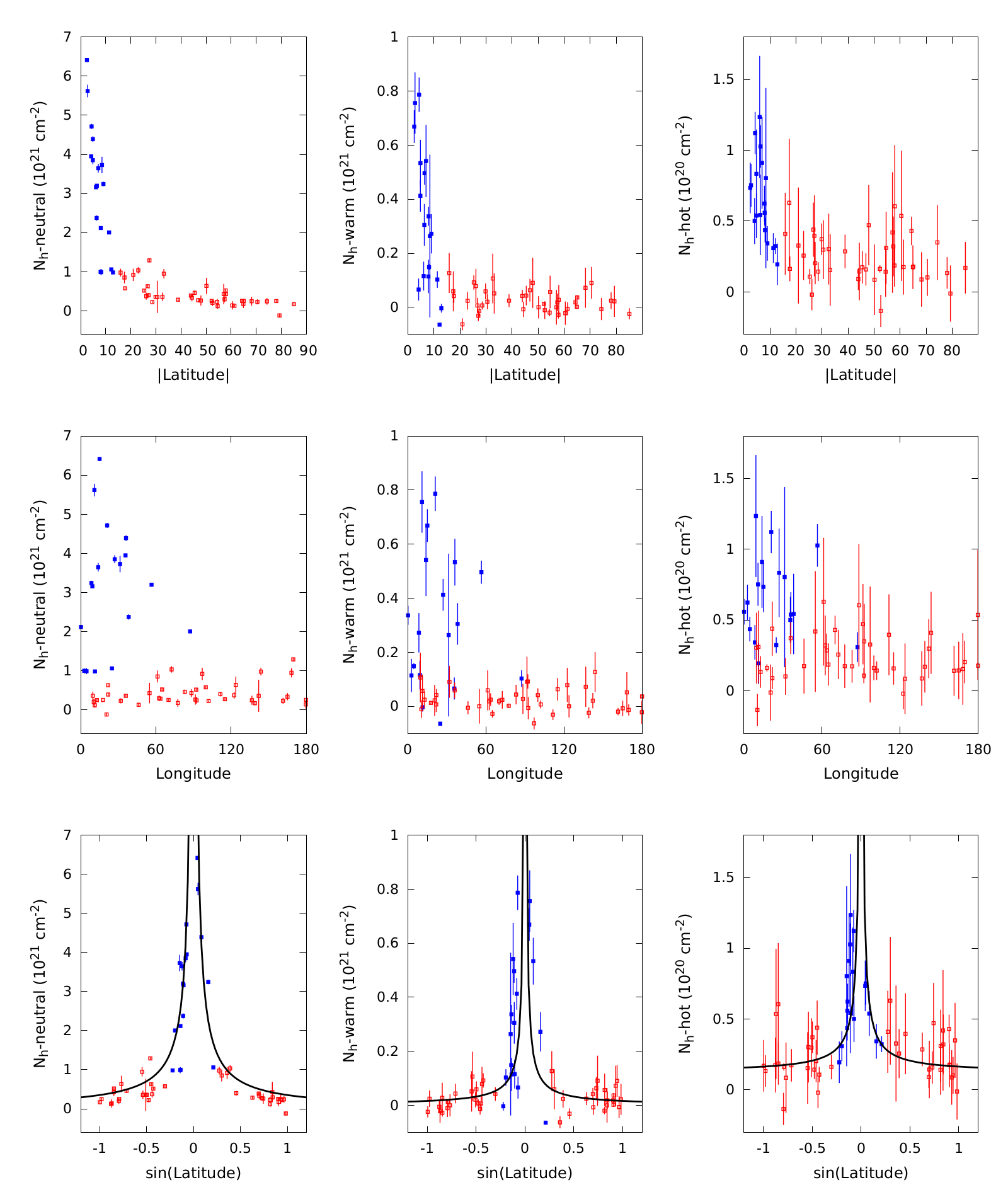}
      \caption{Hydrogen column densities obtained from the {\tt IONeq} fits for the neutral (left panels), warm (middle panels) and hot (right panels) components as function of the galactic latitude (upper panels) and galactic longitude (bottom panels, parametrized as $360-l$ for $l>180$).  Blue filled squares correspond to galactic sources while red empty squares correspond to extragalactic sources. }\label{fig_lat_lon}
   \end{figure*} 
  
Tables~\ref{tab1} and~\ref{tab2} show the $N({\rm H})$ values from the best fits for all sources analyzed. For sources with {\it Chandra} and {\it XMM-Newton} observations, the $N({\rm H})$ values correspond to the mean value between them. Negative column densities, although unphysical, are possible due to statistical fluctuations \citep{mil13}. We include these negative values in our final analysis to avoid possible bias in the sample. For the galactic sources we found typical values of $N({\rm H})_{neutral}\sim 3\times 10^{21}{\rm cm}^{-2}$, $N({\rm H})_{warm}\sim 3 \times 10^{20}{\rm cm}^{-2}$ and $N({\rm H})_{hot}\sim 2 \times 10^{20}cm^{-2}$. For the extragalactic sources we found average values of $N({\rm H})_{neutral}\sim 4 \times 10^{20}{\rm cm}^{-2}$, $N({\rm H})_{warm}\sim 3  \times 10^{19}{\rm cm}^{-2}$ and $N({\rm H})_{hot}\sim 2  \times 10^{19}{\rm cm}^{-2}$.

Figure~\ref{fig_lat_lon} shows column densities of the neutral-warm-hot $N({\rm H})$ components from the {\tt IONeq} model as function of latitude and longitude for all analyzed sources.  Galactic and extragalactic  sources are shown as  blue filled and red empty squares, respectively. The plots show very strong correlation of the column densities with the Galactic latitude and, less prominently, with the longitude.

\begin{figure}[H]
       \begin{center}
\includegraphics[width=8.4cm,height=22cm]{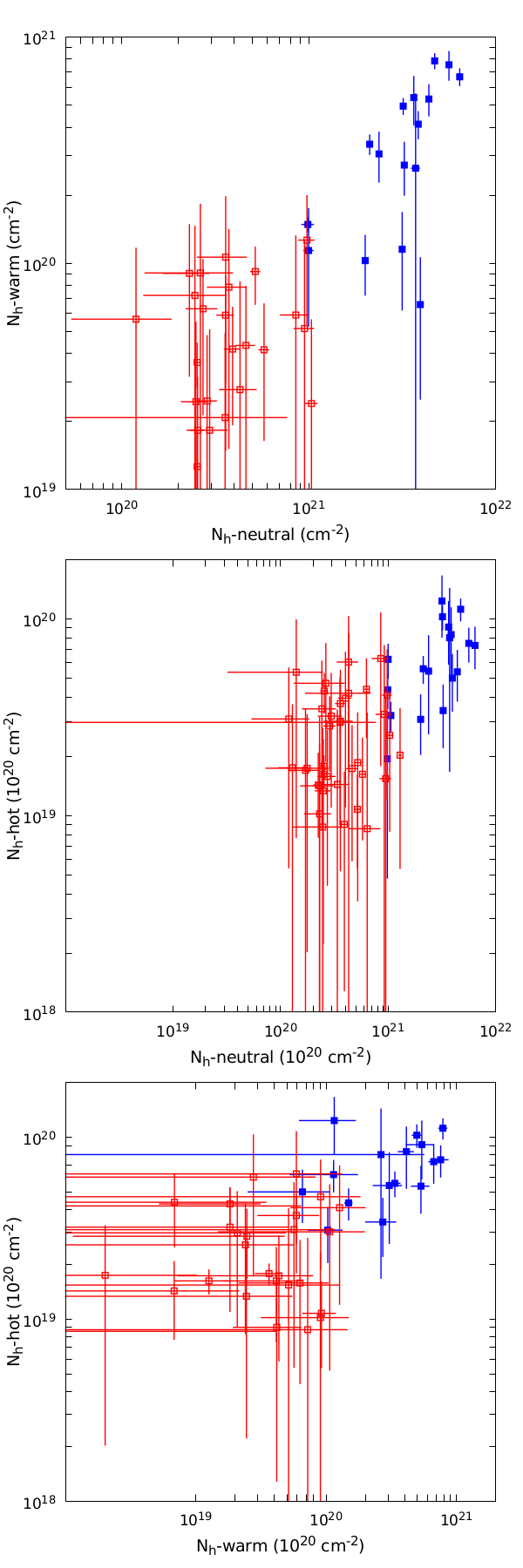}
      \caption{Comparison between hydrogen column densities obtained for the neutral-warm-hot components. Blue filled squares correspond to galactic sources while red empty squares correspond to extragalactic sources.}\label{fig_to_neutral}
      \end{center}
   \end{figure}

Figure~\ref{fig_to_neutral} shows a comparison between the $N({\rm H})$ values for the different ISM components. It is clear that i) the neutral component has the largest column density and ii) there is a positive correlation between column densities of different components. Such good correlation suggest that the contribution from intrinsic (to the sources) absorption is subdominant. 

\begin{figure}
       \begin{center}
\includegraphics[width=8.4cm,height=7cm]{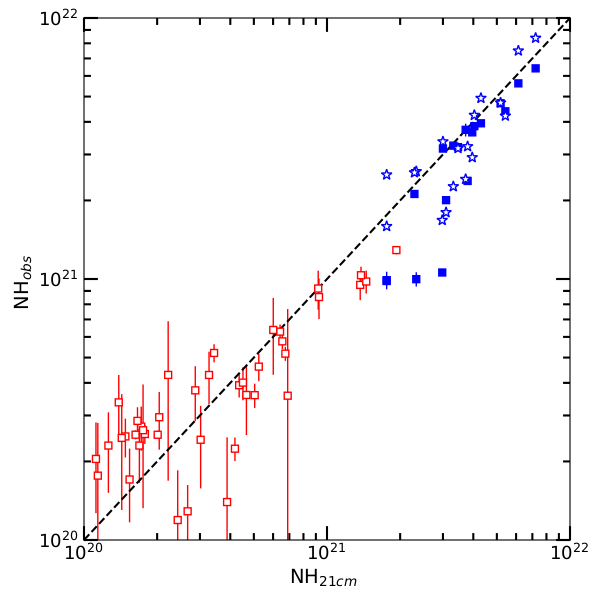}
      \caption{Comparison of the 21~cm measurements from \citet{wil13} and those derived from the {\tt IONeq}(neutral) model.  Galactic sources are indicated with blue filled squares while extragalactic sources are indicated with red empty squares. Blue stars correspond to predicted $N({\rm H})$ values for galactic sources asuming large distances and the density profile described by Equation~\ref{eq:nss}.}\label{fig_21cm}
      \end{center}
   \end{figure}

Figure~\ref{fig_21cm} shows a comparison between 21 cm survey measurements from \citet{wil13}, which include molecular gas, and the column densities derived from the {\tt IONeq}(neutral) model. Galactic and extragalactic sources are indicated with blue filled and red empty squares, respectively. The column densities obtained for galactic sources tend to be underestimated by our model. This is expected given that 21~cm surveys provide $N({\rm H})$ measurements over the entire line of sight while some of the galactic sources are only few kpc away from us. For example, blue stars on Figure~\ref{fig_21cm} show the $N({\rm H})$ value for galactic sources predicted from the density model for the neutral component described by Equation~\ref{eq:nss} and assuming a large distance (See Section~\ref{sec_density} below). Also, it was pointed out by \citet{gat16} that discrepancies between 21~cm survey measurements and values obtained from fitting procedure may arise from intrinsic absorption (e.g. strong stellar winds) or due to the continuum modeling. For the sake of consistency and simplicity we have used a {\tt Bknpower} component to model the continuum for all sources. 

   \begin{figure}
       \begin{center}
\includegraphics[width=8.4cm,height=7cm]{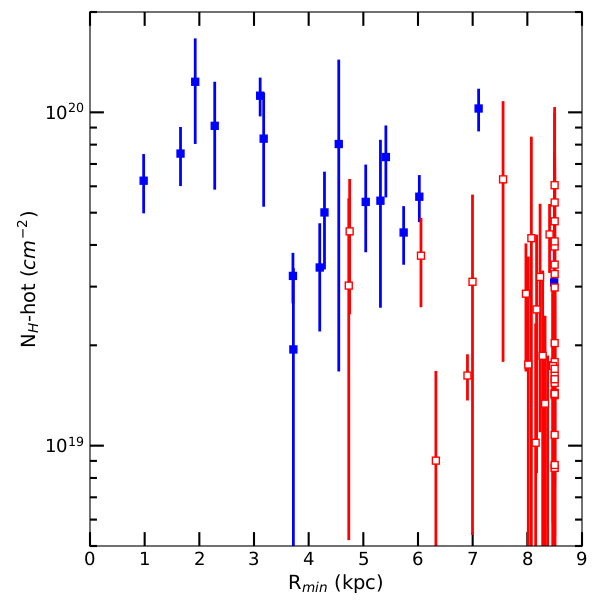}
      \caption{Minimum distance $R_{min}$ from the galactic center to the line of sight between the source and the observer for the X-ray sources included in our sample. Blue filled squares correspond to galactic sources while red empty squares points correspond to extragalactic sources}\label{fig_rmin}
      \end{center}
   \end{figure} 
 
\section{Spatial distribution of the gas}\label{sec_density}
For a given ionic specie $i$, the column density $N_{i}$ depends on its density distribution $n_{i}(x)$ as
\begin{align}
&&N_{i}=\int _{observer}^{source} n_{i}(x)dx,
\end{align}
where $x$ is the distance along the line of sight. This relation can be used to constrain the 3D distribution of the gas.

\begin{figure}
       \begin{center}
\includegraphics[width=7.5cm,height=6cm]{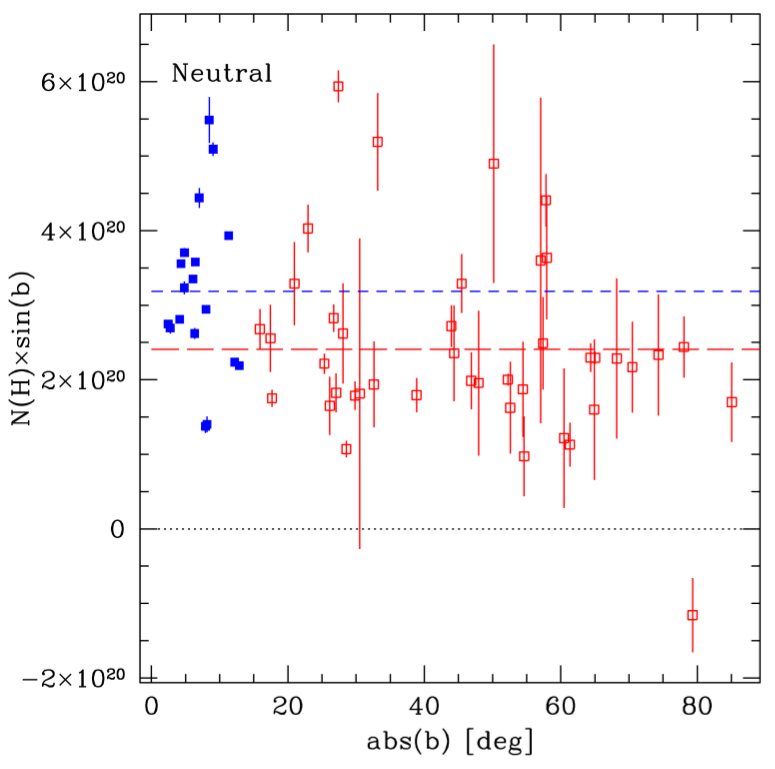}\\
\includegraphics[width=7.5cm,height=6cm]{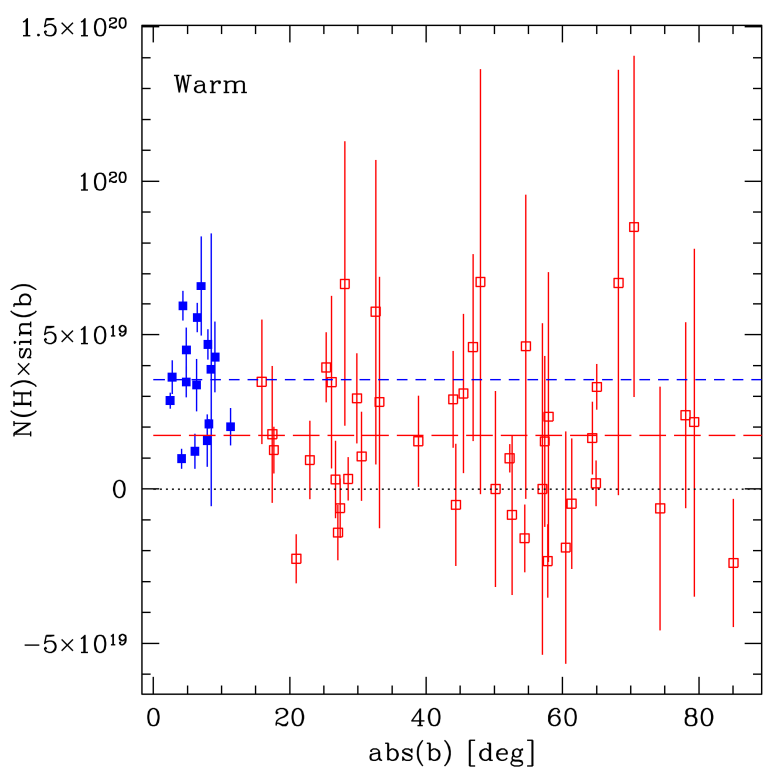}\\
\includegraphics[width=7.5cm,height=6cm]{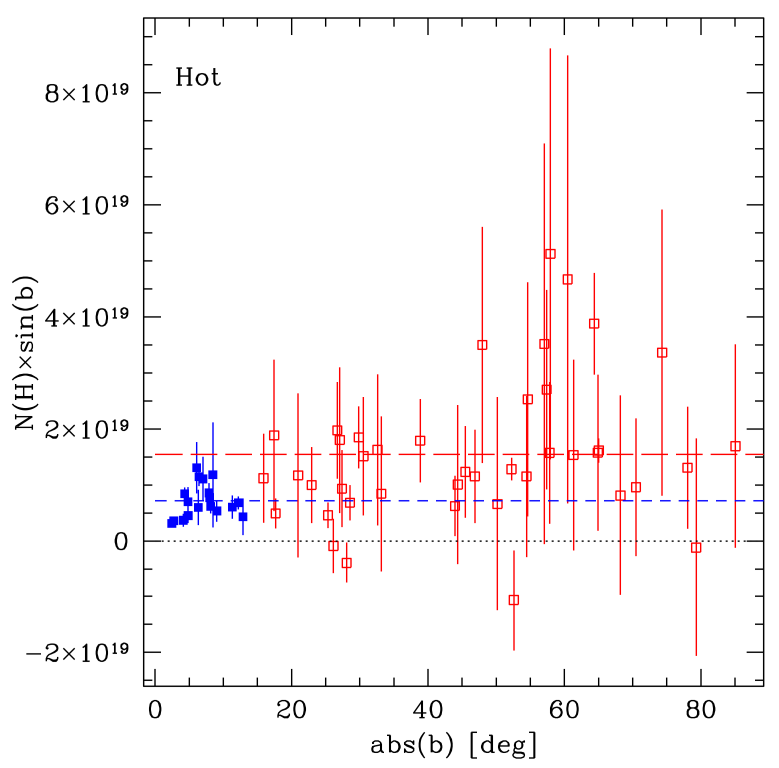}
      \caption{Observed column densities multiplied by the $|\sin b|$ as a function of $|b|$ for the Neutral, Warm and Hot components, respectively. The galactic and extragalactic sources are shown with blue filled squares and red empty squares, respectively. Horizontal lines correspond to the mean value for the galactic (blue line) and extragalactic (red line) sources. In the infinitely thin disk model such product should be the same for all sources. For the Neutral component, the galactic sources (mostly located in the general direction of the Galactic center, have larger values then the extragalactic ones, suggesting that the disk has a finite thickness and the Neutral gas density is larger towards Galactic Center. For the Warm component, roughly similar trend is present. For the hot component, the trend is reversed, suggesting that on top of the disk component, there is an additional extended hot ISM component that contributes to the measured column densities for extragalactic objects.}
      \label{fig:sinb}
      \end{center}
   \end{figure}

As a staring point we consider implications of the most simple possible model, consisting of a very thin disk. In this case, the expected column density for each source is expected to follow a simple law $N(H)=N_{z}/|\sin b|$, where $b$ is the galactic latitude and $N_z$ is the normalization (vertically integrated disk column density from the disk mid-plane to infinity). To this end, we plot in the Figure~\ref{fig:sinb} the values of $N_z=N(H)|\sin{b}|$ as a function of $|b|$. In the infinitely thin disk model the $N_z$ should be the same for all sources. Figure~\ref{fig:sinb} shows that for the Neutral and Warm components, the galactic sources have on average larger values of $N_z$. For the Hot component the trend is in opposite direction. We use this plot as a guide line to build slightly more sophisticated models below.

\subsection{Neutral component}\label{sec:spa_n}
Given strong dependence of Neutral component column density on $b$ (Figure~\ref{fig_lat_lon}) and hints on the difference in $N_z$ for galactic and extragalactic sources (Figure~\ref{fig:sinb}), a disk model with radially declining density seems appropriate. To this end we used the exponential profile model:
\begin{align}
n_{i}(x)=n_{0}\exp(-R/R_{c})\exp(-|z|/z_{c})\label{eq_neutral}
\end{align}
where $n_{0}$ is the normalization, $R_{c}$ is the core radius, $z_{c}$ is the core height and ($R$, $z$) are the Galactocentric cylindrical coordinates, which are related to $l$, $b$ and $x$ as
\begin{align}
R^{2}=x^{2}\cos^{2}(b)-2xR_{sun}\cos(b)\cos(l)+R_{sun}^{2}\\
z=x\sin(b)
\end{align}
where $R_{sun}=8.5$ kpc is the adopted distance from the galactic center to the Sun.  Given that our sample lacks the objects in close vicinity of the Galactic Center, as well as objects closer to the Earth than $\sim$2 kpc, one can expect relatively weak constraints on the thickness of the disk $z_c$ (as long as it is small) and on the central density. 
There are two other issues associated with the analysis of cold component. Firstly, due to the large range of column densities of the neutral component, varying from $\sim 10^{20}$ to $\sim 6\times10^{21}~{\rm cm^{-2}}$, the signal-to-noise ratio also varies strongly, reaching $\sim 100$ for galactic sources at low $b$. Secondly, the above model (see Equation~\ref{eq_neutral}) predicts smooth distribution of the column densities over the sky, while the observed 21-cm maps do show substantial variations of the intensity on different angular scales. The combination of these two factors makes the results of a straightforward fit of the model to the data dubious, since the sources with the highest column densities (or with longest exposure time) completely dominate the fit, while the entire population of sources with low column densities (extragalactic objects) or low exposure time is essentially ignored. As a result, even very sophisticated models have unacceptably large $\chi^2$ (more than 200 per d.o.f.). To mitigate both problems, we have added quadratically  to all measurements errors i) a systematic error of 30\% and ii) a constant error of $6\times10^{19}~{\rm cm^{-2}}$, which is the median value of the error in our sample. With these modifications the impact of galactic and extragalactic sources (as well as sources with different exposure times) is better balanced. Of course, with this approach the best-fitting values and corresponding uncertainties do not have clear statistical meaning. 

Fitting all free parameters yields $z_c\sim 40\;{\rm pc}$ and $R_c \sim 2.5\;{\rm kpc}$. Fixing $z_c$ to $140$~pc \citep[see, e.g.][]{rob03,kal09b} gives comparable quality fit with  $R_c \sim 6\;{\rm kpc}$. The parameters of the latter model are given in Table~\ref{tab_density}. A comparison of the predicted and measured column densities is shown in Figure~\ref{fig:bf}. According to this model the local (to the Earth) vertically integrated column density of the neutral gas is  $N_{z,Neutral}\sim 2\times10^{20}\;{\rm cm^{-2}}$. This number is expected to be more robust than individual parameters of the model.

\subsection{Warm component}
The column density of the warm component is linked to the {\rm O}~{\sc ii}, {\rm O}~{\sc iii}, {\rm Ne}~{\sc ii} and {\rm Ne}~{\sc iii} lines. The first two lines fall close to the much more prominent OI absorption features making the column density measurements less robust. In addition, for several objects the {\rm O}~{\sc ii} $K_\alpha$ line overlaps with the RGS instrumental features \citep[see, e.g.][]{nic16a}. Two such objects (Swift~J1753.5--0127 and HETEJ1900.1--2455) have been excluded from the analysis of the Warm component.  For other objects the data from Chandra observatory have been used. As for the Neutral component the scatter in the data points is larger than the statistical uncertainties, with a typical value of the $\chi^2$ about 5-6 (per d.o.f.).  As in Section \ref{sec:spa_n} we used the same recipe to modify the errors: a systematic error of 30\% and a medial error value ($\sim 3.6\times10^{19}~{\rm cm^{-2}}$) have been quadratically added to the measurement errors. Motivated by Figures \ref{fig_lat_lon} and \ref{fig:sinb} we decided to use the same model as for the Neutral component, letting the normalization be the only free parameter \citep[see][for more sophisticated models]{nic16a}.  The vertically-integrated column density of the warm component is  $N_{z,Warm}\sim 1.8\times10^{19}\;{\rm cm^{-2}}$.

\subsection{Hot component}
For the hot component, we did not assign any additional systematic uncertainties on top of the measurements errors. The behavior of the $N_z$ with $b$ (see Figure~\ref{fig:sinb}) suggests that a single disk model cannot fit both galactic and extragalactic objects. Accordingly, the 3D distribution of the hot gas was modeled as a combination of two functions: a $\beta$-flattened profile, to account for the contribution from the galactic disk and a spherically symmetric profile to account for the possible galactic halo contribution. The total density profile for the hot component is described as 
\begin{align}\label{eq_model_total}
&&n(x)=n_{f}(x)+n_{s}(x),
\end{align}
where 
\begin{align}
&n_f(x)=n_{f,0}\left[1+(R/R_{f})^{2} + (z/z_{f})^{2})\right]^{-3\beta_{f}/2} \\
&n_s(x)= n_{s,0}\left[1+(\sqrt{R^{2}+z^{2}}/R_{s})^{2}\right]^{-3\beta_{s} /2}\label{eq_model}
\end{align}
For our sample we found that the constraints on $R_f$ are weak, once $R_f\gtrsim 5-10$~kpc. In the  analysis below we fix $R_f$ to a large value (150~kpc), emulating  an extended flattened component (essentially a thick disk). 
If we use only the flattened component,  then the fit yields the disk with the vertical scale height $z_f\sim120$~pc and $\beta_f\sim0.46$ with $\chi^2=100.1$ for 57 degrees of freedom (d.o.f.). Adding a spherical component with characteristic radius $R_s=13$~kpc and $\beta_s=1.1$ reduces the  $\chi^2$ to 89.5  (for 54 d.o.f.). A comparable  value of $\chi^2=91.6$  is obtained by choosing very large $R_s=200$ and small $\beta_s\approx 0$. Such version of the spherical component  
is essentially equivalent to the assumption that the gas is distributed over a large region, so that its contribution to the column densities of galactic sources is small, while for extragalactic objects an additional (constant) value of $N(H)$ is added. In this simplified model (with large $R_f$ and $R_s$), the density in the flattened component is simply
\begin{align}\label{eq:nfs}
&&n_f(x)=n_{f,0}\left[1+(z/z_{f})^{2})\right]^{-3\beta_{f}/2},
\end{align}
which is integrated along the line of sight to get column density, while the contribution of the spherical component to the column density is

\begin{equation}\label{eq:nss}
  N(H)_s =N(H)_{s,0}
    \begin{cases}
      1 & (\text{if \textit{d} > 200 kpc})\\
      d/200 & (\text{if \textit{d} < 200 kpc})
    \end{cases}       
\end{equation}

where $d$ is the distance to the source and 200~kpc is an arbitrary chosen (large) size of the spherical component. As one can see from the above expression only the total column density $N(H)_{s,0}$ is an important parameter.
The best-fitting parameters of this simplified model are given in  Table~\ref{tab_density}. The best-fitting scale height of the flattened component is measured to be $z_f\approx350$~pc, albeit with large uncertainties  - the $\chi^2$ increases by less than 1 with respect to the minimum when $z_f$ is in the range between 0.14 and 1.1 kpc. The vertically integrated column densities of two components of this model are comparable: $\sim 6.8\times10^{18}~{\rm cm^{-2}}$ and $\sim 6.3\times10^{18}~{\rm cm^{-2}}$ for the flattened and spherical components, respectively. Note that these are the effective column densities obtained assuming solar abundance of heavy elements. Should the abundance be different, which is plausibly the case for the spherical component, the column density of the corresponding component should be rescaled as  $\displaystyle N({\rm H})=N({\rm H})_{\rm IONeq}\frac{Z_{\odot}}{Z}$.

  \begin{figure}
       \begin{center}
\includegraphics[width=8.3cm,height=7cm]{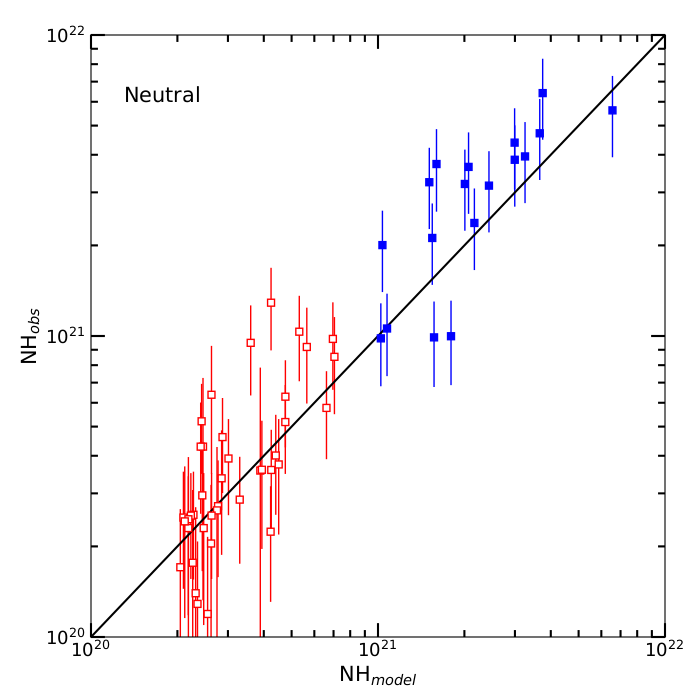}\\
\includegraphics[width=8.3cm,height=7cm]{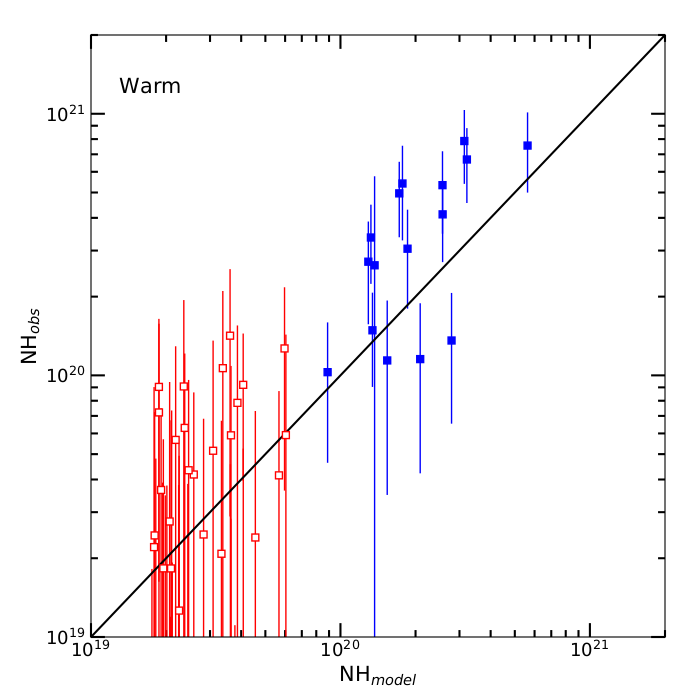}\\
\includegraphics[width=8.3cm,height=7cm]{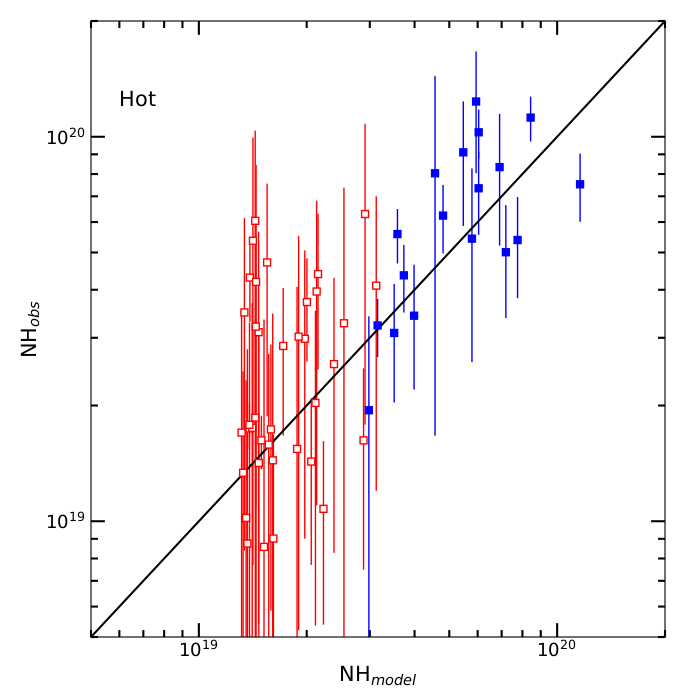}
      \caption{Comparison of the model-predicted and measured column densities for the Neutral, Warm and Hot components. Blue filled squares correspond to galactic sources while red empty squares correspond to extragalactic sources. The errorbars shown correspond to the modified errors for the Neutral and Warm components, and to pure statistical errors for the Hot component. The models broadly follow the trends in the data, but they clearly cannot reproduce full complexity of the measurements.}
      \label{fig:bf}
      \end{center}
   \end{figure}

\begin{table}
\caption{\label{tab_density}Density profiles model parameters.}
\scriptsize
\centering
\begin{tabular}{lllr}
\hline
Model &Parameter  &Units    &Value \\
\hline
\multicolumn{4}{l}{NEUTRAL COMPONENT }\\
Exponential&$n_{0}$&cm$^{-3}$ &2.1   \\ 
&$R_{c}$  &kpc & 5.7 \\
&$z_{c}$ (fixed)&kpc& 0.14\\
\hline
\multicolumn{4}{l}{WARM COMPONENT }\\
Exponential&$n_{0}$&cm$^{-3}$ &$1.8\times 10^{-1}$\\ 
&$R_{c}$ (fixed)&kpc & 5.7\\
&$z_{c}$ (fixed)&kpc&0.14\\
\hline
\multicolumn{4}{l}{HOT COMPONENT }\\
Flattened&$n_{f,0}$&cm$^{-3}$ &$6.7\times 10^{-3}$\\ 
&$z_{f}$ &kpc&0.36\\
&$\beta_{f}$ &&1.1\\
Spherical &$N(H)_{s,0}$&cm$^{-2}$ &$6.3\times 10^{18}$\\ 
Chi-square&$\chi^{2}$/d.o.f&&91.6/56\\
\hline
\end{tabular}
\end{table}

A comparison of the predicted and measured column densities is summarized in Figure~\ref{fig:bf}. While global trends are captured by the models, the data certainly show more complicated behavior than the models described in this section.  We reiterate here that above models do not necessarily provide a comprehensive description of the ISM distribution in the entire Galaxy, given that our sample probes only part of the Galaxy volume. For instance, none of the sources has line of sights passing at a 3D distance of less than $\sim1$ kpc from the Galactic center, as illustrated in Figure~\ref{fig_rmin}. From this point of view, our modeling of the density distribution is most sensitive to the local environment near the Sun, while the behavior of the gas within few kpc from the Galactic Center is not constrained.  Often, LMXB spectra near the galactic center are so heavily affected by the neutral gas absorption that it is difficult to obtain reliable column densities for the hot component (e.g. they have few counts in the oxygen edge absorption region). Different ions, although less abundant, such as  {\rm Si} may be used to trace this component but such an analysis is beyond the scope of this paper.  

\subsection{Local ISM fractions}
As is mentioned before, the vertically integrated column densities $N_{z}$ at the location of the Sun are expected to be more robust than the individual parameters of the full 3D density distribution models. The values of $N_{z}$ obtained for each ISM phase are given in the Table~\ref{tab_simple}. 
The comparison of these values suggests the following mass fractions for these three ISM phases in the Galactic disk: neutral $\sim~89\%$, warm $\sim 8\%$ and hot $\sim 3\%$ components. Of course, these values reflect local (at the location of the Sun) properties. These results broadly agree with previous findings. \citet{pio07} performed numerical simulations including the effects of atomic heating-cooling, magnetic fields, turbulence and vertical gravity, finding that up to $\sim~91\%$ of the ISM mass correspond to the neutral component. \citet{yao06} analyzed the multiphase ISM components using X-ray high-resolution spectra of the LMXB~4U~1820-303 concluding that that the neutral component accounts for $\sim~86\%$ of the total ISM mass along this line of sight. \citet{pin13}, in their analysis of the ISM using 9 LMXB high-resolution X-ray spectra, found that $\sim~90\%$, $\sim~8\%$ and $\sim~2\%$  of the total mass correspond to neutral, warm and hot components, respectively. Finally, \citet{ber08} estimated a degree of ionization of $14\%$ by deriving gas densities from dispersion measures and {\rm H}~{\sc i} column densities towards 393 stars and 191 pulsars. It is important to note that their sample include high-latitude sources, which can lead to a higher fraction of the ionized gas.

\begin{table}
\caption{\label{tab_simple}Vertically integrated local column densities $N_{z}$.}
\small
\centering
\begin{tabular}{lccc}
\hline
Component&$N_{z}$&$N_{s,0}$\\
 &(${\rm cm^{-2}}$)&(${\rm cm^{-2}}$)\\
\hline
Neutral&$2.0$ $\times10^{20}$ &--\\
Warm&$1.8$ $\times10^{19}$ &--\\
Hot&$6.8$ $\times10^{18}$ &$6.3$ $\times10^{18}$\\
\hline
\end{tabular}
\end{table}

\section{Comparison with previous works}\label{sec_previous}
\subsection{Neutral and warm components}\label{sec_neutral_warm}
For the neutral and warm components, our column densities are within $40\%$ of those obtained by \citet{pin13} for their LMXBs included in our sample. \citet{pin13} used a collisionally-ionized equilibrium model included in the {\sc spex} analysis package, although it is important to note that \citet{pin13} assumes that the neutral component is at rest. Also, the abundances in \citet{pin13} vary for each source, while in our case we are assuming solar abundances. Finally, the atomic data (i.e. photoabsorption cross-sections) used by \citet{pin13} do not include Auger-damping, a crucial effect required to model accurately atomic absorption edges \citep{gar05,gat15}.  

Our results for the neutral component are within $35\%$ of the column densities obtained by \citet{gat16} for those LMXBs that are included in our sample. However, for the warm component we found an average of $\approx 2.5$ times their column densities. In this sense, it is worth mentioning that \citet{gat16} used the {\tt ISMabs} X-ray absorption model which does not include broadening effects and the column densities are fitted independently for each ion, without considering ionization equilibrium conditions. 

We obtained column densities within $25\%$ of those obtained by \citet{nic16a}, who analyzed {\rm O}~{\sc i} K$\alpha$ and {\rm O}~{\sc ii} K$\alpha$, K$\beta$ absorption lines and derived column densities using the curve of growth method. We have computed curve of growths for {\rm O}~{\sc i} and {\rm O}~{\sc ii} ions using \citet{nic16a} equivalent widths and we found that, in the case of saturated lines, the column densities can be $\sim 15\%$ higher than those quoted by \citet{nic16a} when including Auger radiative widths. 

\subsection{Hot component}\label{sec_hot}
A similar analysis of the hot gas component associated with the Milky Way has been performed by \citet{mil13,mil15,nic16c}. Main differences between our procedure and their results include (1) \citet{mil13} assumed a turbulent velocity of $b=150$ km/s, consistent with the sound speed of hydrogen while our Doppler parameter was obtained directly from the spectra; (2) our sample includes both {\it Chandra} and {\it XMM-Newton} spectra leading to a better constraints on the column densities; (3) our fitting procedure consider a broad wavelength region (11--24\AA), which include absorption lines from multiple ions. For example, we are modeling at same time {\rm O}~{\sc vii} and {\rm Ne}~{\sc ix} K$\alpha$ transitions for the LMXBs spectra and if we exclude the Ne absorption edge region the column density $N({\rm H})_{warm}$ can differ by a factor of $\sim 3$ in some cases; (4) our density distribution model does not include offsets of the distribution from the Galaxy's center or plane, as implemented in \citet{nic16a} and (5) we are assuming solar abundance ratios and constant metallicities along all line of sights. In this sense, although our column densities can be rescaled to different solar standard abundances, we did not included abundance gradients in this analysis.

\section{Caveats and limitations}\label{sec_cave}
The main caveats in our analysis are as following.
\begin{enumerate}
\item Some of the extragalactic sources included in our sample are know to have intrinsic absorption. For instance, due to the presence of warm absorbers (e.g. NGC~5548, Mrk~279, Mrk~290) or ultra-fast outflow signatures \citep[e.g. IRAS13349+2438,][]{par17}. We did tests on selected sources by modeling their intrinsic absorber together with our baseline ISM model. We found that there is no strong impact on the ISM column densities obtained with the {\tt IONeq} model. We intend to carry out in the near future a detailed analysis of intrinsic X-ray absorption features in extragalactic sources.
\item We assume fixed (solar) abundance ratios when deriving $v_{turb}$. Different abundance ratios will affect the broadening velocity obtained from the fits. Also we assume that the broadening effect is independent on the line-of-sight direction.
\item The lack of dense coverage over sky/distances implies that our ability to place comprehensive constraints on the density distribution is limited (e.g. the absence of sources near the Galactic center).
\item The systematic errors are important, in particular those caused by homogeneity in the true column density distribution which is not captured by 
our models. This makes the estimates of the fit uncertainties for the cold and warm components useless. A large sky coverage is required to address this issue.
\item We assume CIE, deferring the study of photoionization effects for subsequent studies.
\end{enumerate}
\section{Conclusions}\label{sec_con}
High-resolution X-ray spectroscopy is a powerful technique to study the physical properties and spatial distribution of the  absorbing medium in the foreground to bright X-ray sources, such as,  AGNs or galactic binaries. The examples of absorbing medium range from the gas local to the source e.g., outflows from an AGN, to the ISM in the Milky Way. The absorbing properties of the absorbing medium depend on variety of parameters, including abundance of heavy elements, ionization state of the medium, turbulent velocities. To this end, we have developed a model that can account for these parameters, having in mind most common astrophysical applications. In this paper we do some basic tests of the model in application to the Milky Way ISM.  In this section we briefly summarize our findings.

\begin{enumerate}
\item We have developed a new X-ray absorption model, called {\tt IONeq}, which computes the optical depth $\tau(E)$ taking into account ions of all astrophysically-abundant elements, assuming (collisional and photo) ionization equilibrium. The parameters of the model are the gas temperature $_{e}$, ionization parameter $\log(\xi)$, hydrogen column density $N({\rm H})$ of the absorber, redshift ($z$) and turbulent broadening velocity ($v_{turb}$). The model assumes standard solar abundances from \citet{gre98}. The model is publicly available at \url{https://heasarc.gsfc.nasa.gov/docs/xanadu/xspec/models/ioneq.html}.

\item While the model includes photoionization equilibrium, here we assume pure collisional ionization equilibrium and measure the absorption column densities in the Milky Way. The model was applied to 18 galactic (LMXBs) and 42 extragalactic sources (mainly Blazars). The velocity broadening $v_{turb}$ was estimated from fitting the spectra of the two brightest sources in our sample (Cygnus~X-2 and PKS~2155-304) and then used for all other objects in the sample. For each source we fit a broken-power-law continuum model modified by the ISM absorption, provided by three distinct components: neutral, warm and hot. The ``neutral'' component, with a characteristic temperature below $\log($T$_{e})\lesssim 4$, is primarily traced by {\rm O}~{\sc i} K$\alpha$, the {\rm Ne}~{\sc i} absorption edge and metallic iron absorption features. The ``warm'' component, with a characteristic temperature $\log($T$_{e})=4.7$, is  traced by {\rm O}~{\sc ii}, {\rm O}~{\sc iii}, {\rm Ne}~{\sc ii} and {\rm Ne}~{\sc iii} K$\alpha$ absorption lines. A hot component, with temperature $\log($T$_{e})=6.30$, can be identified by {\rm O}~{\sc vii}, {\rm O}~{\sc viii}, {\rm Ne}~{\sc ix} K$\alpha$ and {\rm O}~{\sc vii}  K$\beta$ absorption lines in the spectra. For each of these components we have determined an equivalent hydrogen column density needed to explain the observed absorption features, assuming solar abundances $Z_{\odot}$. For different abundance $Z$, the column densities obtained from {\tt IONeq} can be easily rescaled as $\displaystyle N({\rm H})=N({\rm H})_{\rm IONeq}\frac{Z_{\odot}}{Z}$.

\item For the objects with the largest column densities, the correlation between column densities for all three ISM phases is evident in the data, as well as the correlation with the galactic latitude. These results show that  (at least for these objects) the measured absorption features are due to the ISM, rather than being intrinsic to the studied objects. 

\item The derived column densities were used to constrain the ISM density distribution models in the Galaxy. For the neutral and warm components any smooth model gives unacceptably large $\chi^2$, at least partly (and, possibly, predominantly) due to inhomogeneities of the true column density distribution over the sky. For that reason we decided to use a a minimalistic density distribution model $n=n_{0}\exp(-R/R_{c})\exp(-|z|/z_{c})$, where $R_c\sim 5.7\;{\rm kpc}$ and $z_c\sim 14\;{\rm pc}$ for both components. With these parameters, the disk midplane densities at the Galactic Center are 2.1 and $0.18\;{cm^{-3}}$ for the neutral and warm components, respectively. The local (to the disk at the Sun galactocentric radius $R\sim 8\;{\rm kpc}$) vertically integrated column densities  are $N_{z,\rm Neutral}=2\times10^{20}$ and $N_{z, \rm Warm}=1.8\times10^{19}\;{\rm cm^{-2}}$. These quantities are expected to be more robust than other parameters of the model separately. 

\item For the hot component,  the disk model is clearly insufficient and we adopted a model that 
includes both the contributions of a disk and  a halo. The halo component is characterized by its total column density $N_{\rm halo}=$ 6.3 $\times 10^{18} {\rm cm^{-2}}$, which is comparable to the vertically integrated disk column 
density $N_{z, \rm disk}=$ 6.8 $\times 10^{18} {\rm cm^{-2}}$. The constraints on the radial distribution of the halo component are weak. 

\item The comparison of the  vertically integrated column densities suggests the following mass fractions for these three ISM phases in the Galactic disk: neutral $\sim~89\%$, warm $\sim 8\%$ and hot $\sim 3\%$ components, respectively. If the comparison is done using the 21~cm measurements for the neutral component, the mass fractions are similar to these values. 

\item The absence of sources for which we can measure O and Ne lines near the galactic center requires that some assumptions about the size and thickness of the galactic disk be made. Observations focused on other ions, such as {\rm Si}, may provide insights on the inner part of the galactic density profile.

\end{enumerate}

  \appendix
\label{sec_apx}

\section{Spectra of individual source and best-fitting models}\label{sec_apx_fit}
\begin{figure*}
        \begin{center}
\includegraphics[scale=0.3]{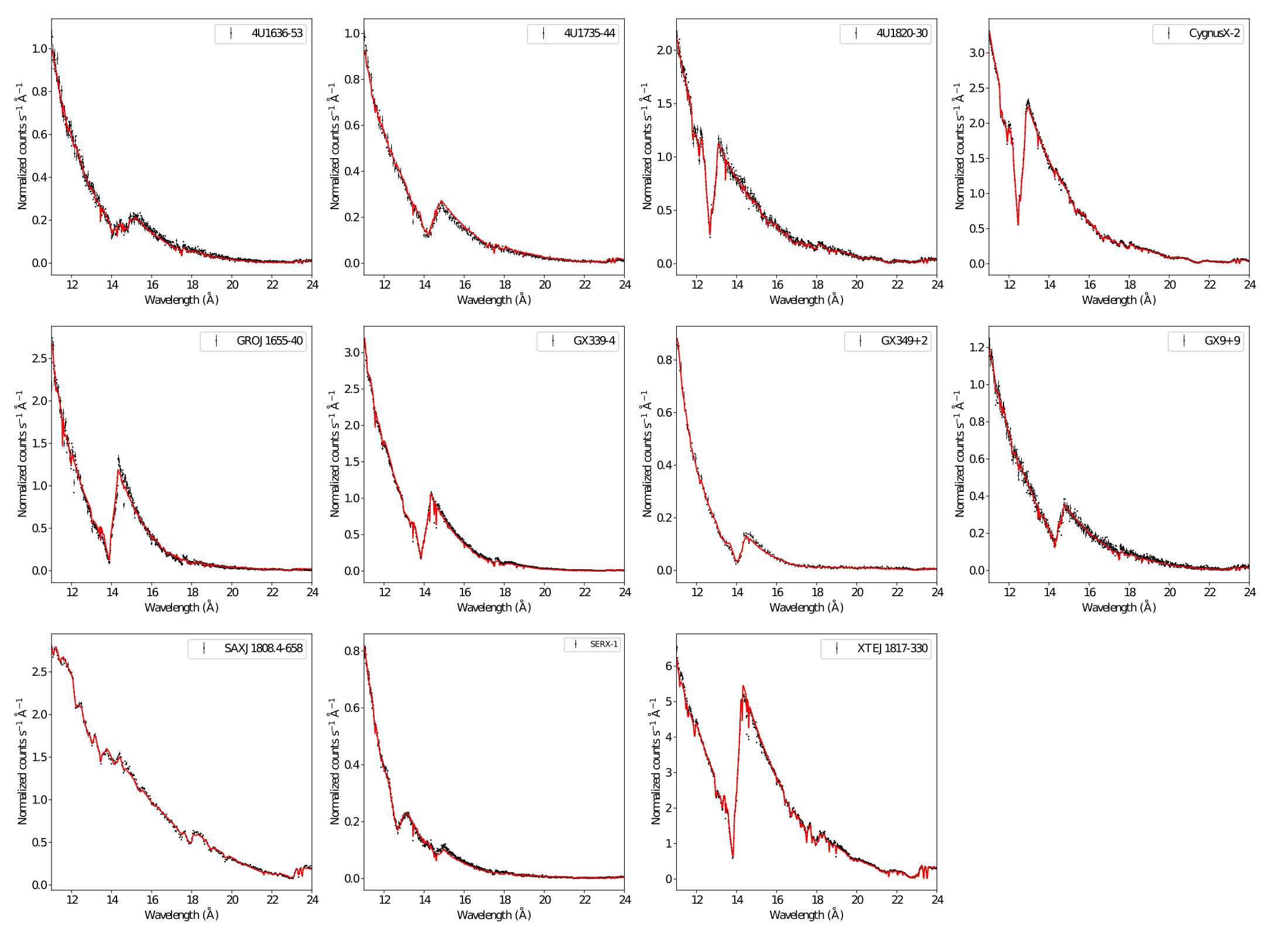}
      \caption{{\tt IONeq} best fits for LMXBs {\it Chandra} MEG spectra. Observations are combined for illustrative purposes. }\label{fig_lmxb_chandra}
      \end{center}
   \end{figure*}

\begin{figure*}
        \begin{center}
\includegraphics[scale=0.3]{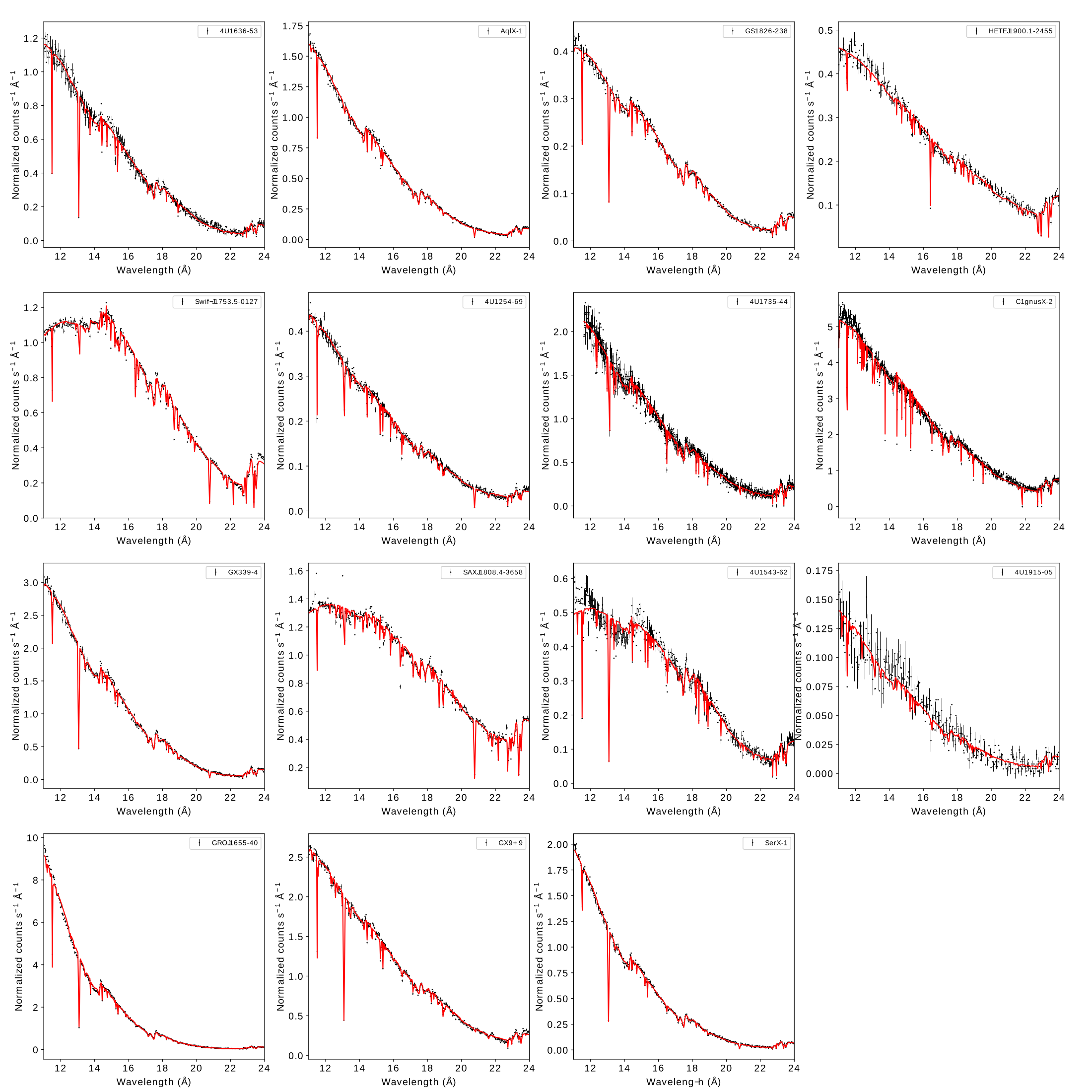}
      \caption{{\tt IONeq} best fits for LMXBs {\it XMM-Newton} spectra. Observations are combined for illustrative purposes.}\label{fig_lmxb_xmm}
      \end{center}
   \end{figure*}

\begin{figure*}
        \begin{center}
\includegraphics[scale=0.25]{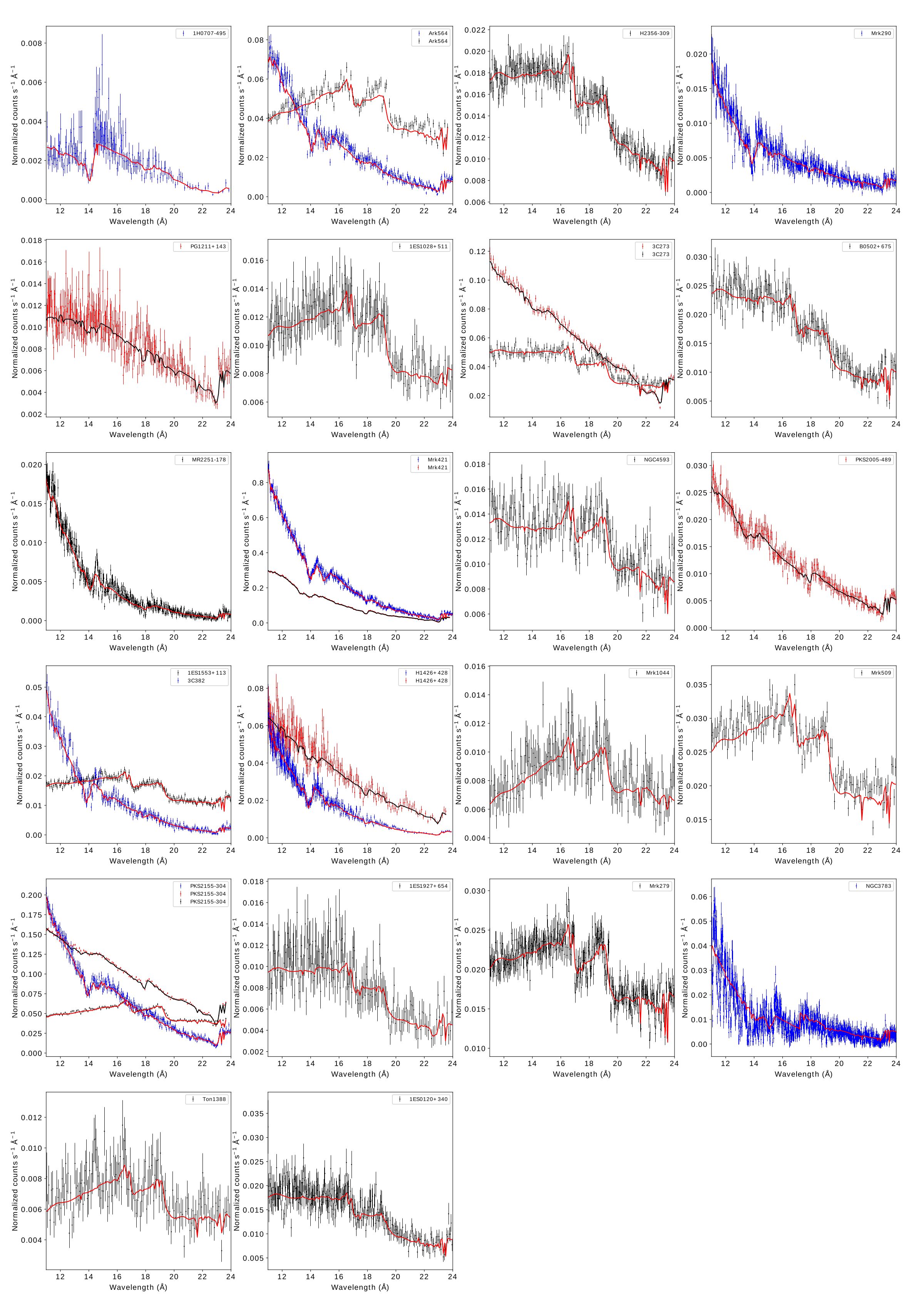}
      \caption{{\tt IONeq} best fits for extragalactic {\it Chandra} LETG-HRC spectra (black points), LETG-ACIS spectra (red points) and MEG spectra (blue points). Observations are combined for illustrative purposes. }\label{fig_ex_chandra}
      \end{center}
   \end{figure*}

\begin{figure*}
        \begin{center}
\includegraphics[scale=0.22]{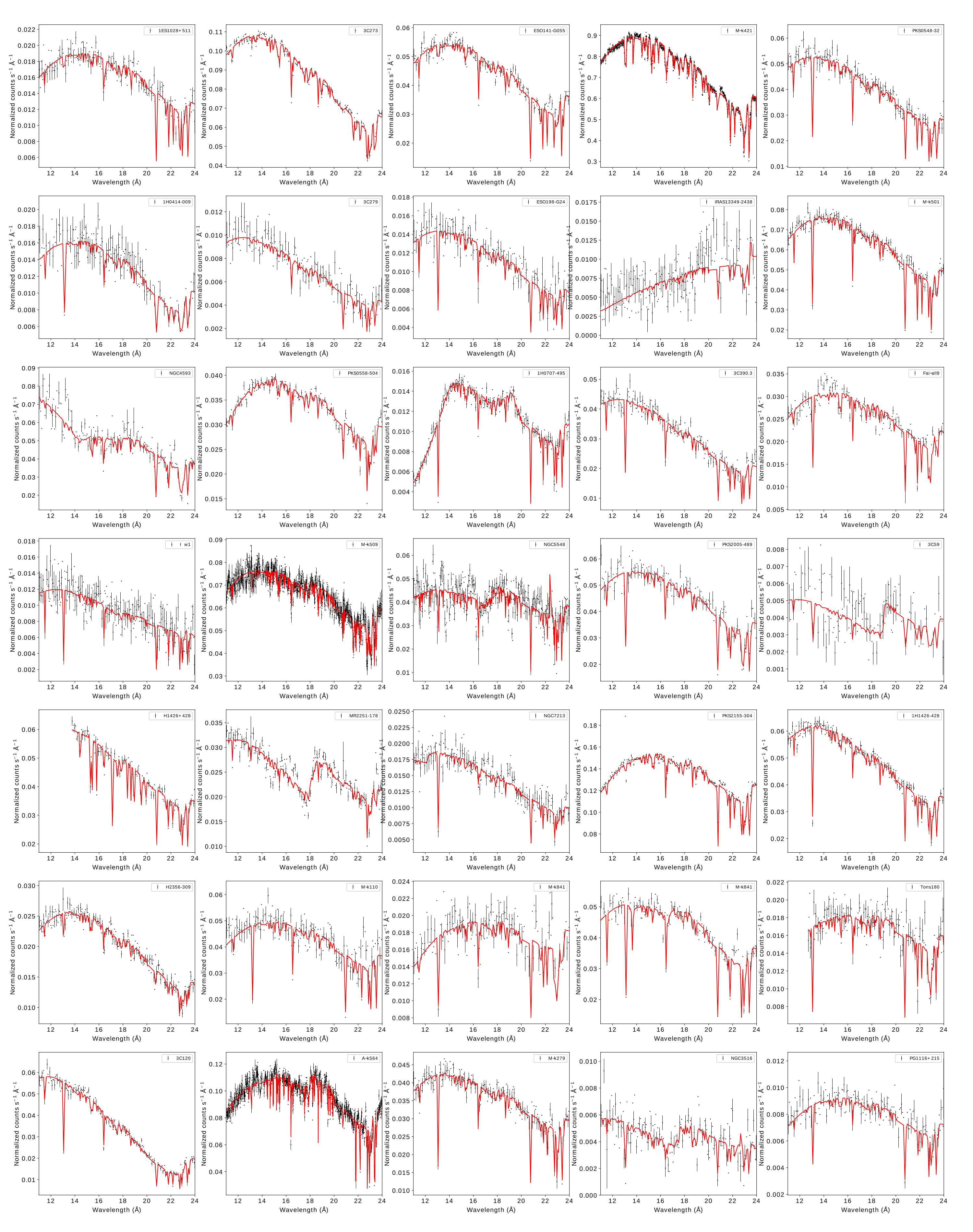}
      \caption{{\tt IONeq} best fits for extragalactic {\it XMM-Newton} spectra. Observations are combined for illustrative purposes.}\label{fig_ex_xmm}
      \end{center}
   \end{figure*}

Figures~\ref{fig_lmxb_chandra} and~\ref{fig_lmxb_xmm} show the best-fitting models obtained for the {\it Chandra} and {\it XMM-Newton} LMXBs samples, respectively. For each source the spectra measured in individual observations were fitted simultaneously. In the plots these spectra are combined for illustrative purposes. The best-fitting models are shown with the solid red line. In the case of {\it Chandra} all spectra were taken with the Medium Energy Grating (MEG) on board of the High-Energy Transmission Grating (HETG) instrument. Figure~\ref{fig_ex_chandra}  shows the best-fitting models obtained for {\it Chandra} extragalactic sources. Black points correspond to observations obtained with the Low-Energy Transmission Grating (LETG) in combination with the High-Resolution Camera (HRC). Red points correspond to observations obtained with the LETG in combination with the Advanced CCD Imaging Spectrometer (ACIS). Finally, blue points correspond to MEG observations. Figure~\ref{fig_ex_xmm} shows the best-fit obtained for {\it XMM-Newton} extragalactic sources. For all sources we obtained a good fit with $\chi^{2}/d.o.f<1.10$

\bibliographystyle{mnras}

\end{document}